\documentclass[pdflatex,sn-mathphys-num]{sn-jnl}% Math and Physical Sciences Numbered Reference Style
\usepackage{graphicx}%
\usepackage{multirow}%
\usepackage{amsmath,amssymb,amsfonts}%
\usepackage{amsthm}%
\usepackage{mathrsfs}%
\usepackage[title]{appendix}%
\usepackage{xcolor}%
\usepackage{textcomp}%
\usepackage{manyfoot}%
\usepackage{booktabs}%
\usepackage{tabularx}%
\usepackage{algorithm}%
\usepackage{algorithmicx}%
\usepackage{algpseudocode}%
\usepackage{listings}%
\usepackage{subcaption}
\usepackage{hyperref}
\usepackage[numbers,sort&compress]{natbib}
\theoremstyle{thmstyleone}%
\theoremstyle{thmstyletwo}%
\theoremstyle{thmstylethree}%
\begin{document}

\title[Article Title]{Constraint-Preserving QAOA for Personnel Rostering: Coverage-Preserving and Guarded-XY Mixer Constructions}
\author*[1]{\fnm{Aruna} \sur{Gupta}}\email{arunagupta169@gmail.com}
\author[1]{\fnm{S. R.} \sur{Hassan}}\email{shassan@imsc.res.in}
\affil[1]{\orgdiv{Department of Theoretical Physics}, \orgname{The Institute of Mathematical Sciences, A CI of Homi Bhabha National Institute}, \orgaddress{\street{CIT Campus, Taramani}, \city{Chennai}, \postcode{600113}, \state{Tamil Nadu}, \country{India}}}

\abstract{The Quantum Approximate Optimization Algorithm (QAOA) is a promising framework for combinatorial optimization, but constrained problems are commonly handled using energetic penalty terms that require calibration and allow infeasible configurations to remain dynamically accessible. We develop a constraint-preserving QAOA framework for personnel rostering in which hard scheduling constraints are embedded directly into the mixer Hamiltonian. Using a binary rostering model with daily coverage and no-consecutive-duty constraints, we formulate the dynamics from a transition-graph perspective and introduce a guarded-XY mixer that confines the evolution to the fully feasible scheduling subspace. We further distinguish feasibility preservation from feasible-transition design and propose a tight-pattern extension that introduces collective feasible exchanges in saturated workload segments where local guarded exchanges alone are insufficient. Exact statevector simulations demonstrate that, compared with Penalty-X and Coverage-XY formulations under both expectation-value and Conditional Value-at-Risk optimization, the proposed approach eliminates hard-constraint penalty calibration, guarantees feasible evolution by construction, and consistently yields higher-quality output distributions with stronger concentration on optimal feasible schedules. To the best of our knowledge, this is the first constraint-preserving QAOA formulation for personnel rostering, and the transition-graph framework is readily applicable to a broad class of constrained quantum optimization problems.}
\maketitle
\section{Introduction}
Many important optimization problems involve hard operational constraints in addition to an objective function. Examples include personnel scheduling, resource allocation, routing, portfolio optimization, timetabling, and assignment problems. Although integer programming, constraint programming, local search, and metaheuristic algorithms have achieved considerable success, many constrained optimization problems remain computationally challenging because the search space grows exponentially with system size. These challenges have motivated increasing interest in quantum optimization algorithms, particularly hybrid quantum--classical methods designed for noisy intermediate-scale quantum (NISQ) devices \cite{Preskill2018NISQ}.

Among gate-based quantum algorithms, the Quantum Approximate Optimization Algorithm (QAOA) has become one of the most widely studied approaches for combinatorial optimization \cite{Farhi2014QAOA}. QAOA alternates between a phase Hamiltonian encoding the optimization objective and a mixer Hamiltonian that explores the computational basis. In the standard formulation, the mixer is a transverse field acting independently on each qubit. While this produces simple and highly connected dynamics, it does not preserve problem-specific constraints. Consequently, constrained optimization problems are usually formulated by adding energetic penalty terms to the phase Hamiltonian so that infeasible configurations become energetically unfavorable. 

Penalty-based formulations are simple and general, but they have well-known drawbacks. Their performance may depend sensitively on the choice of penalty coefficients, large penalties can distort the optimization landscape, and infeasible configurations remain dynamically accessible throughout the variational evolution. As a result, quantum amplitude can accumulate in infeasible regions of the Hilbert space even when those states are energetically penalized. These limitations have motivated constraint-preserving variants of QAOA. Within the Quantum Alternating Operator Ansatz framework, the mixer Hamiltonian can be designed so that the evolution remains confined to a feasible subspace of the computational basis \cite{Hadfield2019QAOAtoQAOAplus}. A well-known example is the XY mixer, which conserves excitation number and therefore naturally preserves fixed-cardinality constraints \cite{Wang2020XYMixers}. Such constructions move part of the constraint handling from the phase Hamiltonian to the variational dynamics.

Personnel rostering provides a natural benchmark for studying this idea because it combines multiple interacting hard constraints while remaining representative of a broad class of scheduling problems. Classical nurse-rostering has been studied extensively using integer programming, constraint programming, and metaheuristic methods, while quantum formulations have mainly focused on QUBO and Ising models for annealing devices \cite{Ikeda2019NurseQA}. In contrast, gate-based QAOA formulations remain comparatively unexplored. To the best of our knowledge, a systematic constraint-preserving QAOA framework for personnel rostering has not previously been reported.

In this work we consider a simplified binary rostering model in which each worker is either assigned or not assigned on each day. The model contains two hard constraints: a daily coverage requirement and a no-consecutive-duty constraint. Although considerably simpler than industrial nurse-rostering models, which include multiple shift types, worker preferences, fairness criteria, skill compatibility, and labour regulations, it captures the essential interplay between staffing requirements and temporal feasibility \cite{Burke2004NurseRostering, Miller1976NurseScheduling, Warner1976NursePreference}.
The daily coverage constraint possesses a natural particle-conservation structure and can therefore be enforced by same-day XY exchanges, leading to the conventional Coverage-XY mixer \cite{Hadfield2019QAOAtoQAOAplus, Wang2020XYMixers}. However, preserving coverage alone is insufficient because an exchange on one day may still violate the no-consecutive-duty constraint. Consequently, the evolution remains confined only to a partially feasible subspace and still requires energetic penalties.

The central contribution of this work is a guarded transition operator that embeds both hard scheduling constraints directly into the mixer Hamiltonian. Instead of generating all elementary exchanges and suppressing infeasible schedules energetically, the guarded mixer removes infeasible transitions from the dynamics itself. This follows the broader constraint-preserving mixer philosophy developed for QAOA~\cite{Fuchs2022ConstraintPreservingMixers}. The resulting evolution remains entirely within the fully feasible scheduling subspace, eliminating the need for hard-constraint penalty calibration.

Viewing mixers through their induced transition graphs further shows that exact feasibility preservation alone is not sufficient for efficient optimization. Under certain workload patterns, the feasible-transition structure induced by local guarded exchanges can become severely restricted. This motivates our second contribution: the separation of feasibility preservation and feasible-transition design as two independent design principles for constraint-preserving QAOA \cite{Cook2020QAOAMaxKVertexCover}. To improve feasible-sector exploration while preserving exact feasibility, we introduce a tight-pattern transition operator that adds collective alternating-pattern moves in saturated workload segments. These constructions naturally define the hierarchy
\begin{equation}
\mathcal H
\supset
F_{\mathrm{cov}}
\supset
F_{\mathrm{full}},
\end{equation}
corresponding to unconstrained, coverage-preserving, and fully constraint-preserving dynamics. Using exact statevector simulations, we compare the conventional Penalty-X, Coverage-XY, and proposed Guarded-XY formulations under both expectation-value and CVaR optimization using common benchmark instances and performance metrics \cite{Barkoutsos2020CVaR}. The results show that embedding both hard constraints directly into the mixer improves feasible-state concentration and output-distribution quality while eliminating the need for hard-constraint penalty calibration.

Beyond personnel rostering, the framework developed here is applicable to a broad class of constrained combinatorial optimization problems in which feasibility can be characterized through local transition rules. More generally, it suggests that the design of QAOA mixers should be viewed not merely as the choice of a Hamiltonian, but as the construction of a transition graph whose geometry determines both the accessible search space and the efficiency with which it is explored.

The remainder of this paper is organized as follows. Section~2 formulates the personnel-rostering problem within the QAOA framework and develops the constraint-preserving mixer construction, including the Coverage-XY, Guarded-XY, and tight-pattern transition operators. Section~3 describes the benchmark instances, simulation protocol, penalty calibration, Hamiltonian rescaling, initialization strategies, and performance metrics used throughout the numerical study. Section~4 presents and discusses the numerical results, comparing the Penalty-X, Coverage-XY, and Guarded-XY formulations under expectation-value and CVaR optimization, and analyzes the effects of constraint preservation, feasible-state connectivity, workload structure, initialization, and problem size. Finally, Section~5 summarizes the main conclusions and discusses possible extensions of the proposed framework to more general constrained quantum optimization problems.

\section{Mathematical Formulation}
\subsection{Rostering Problem, Feasible Sectors, and Standard QAOA}
\begin{figure}[t]
    \centering
\includegraphics[width=0.65\linewidth]{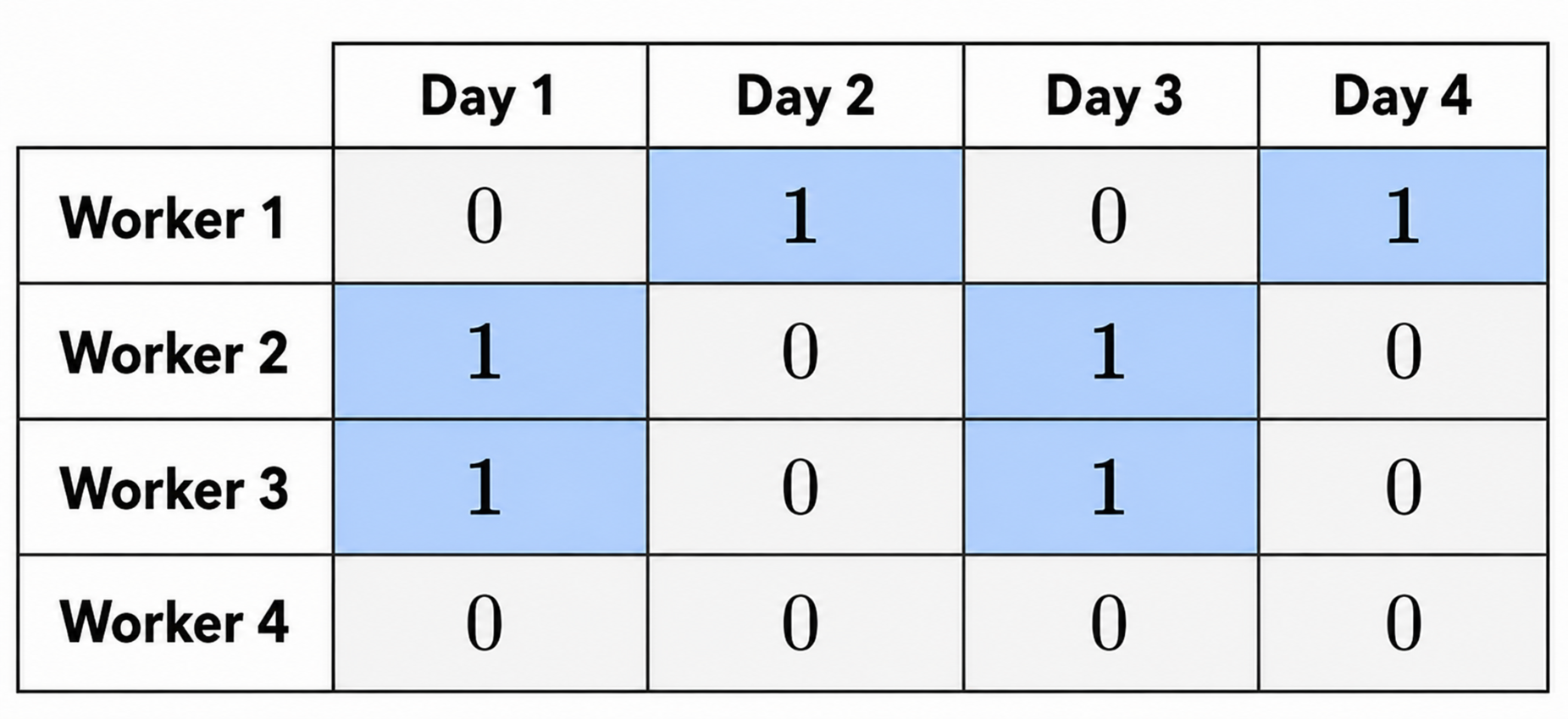}
    \caption{Schematic worker--day binary encoding of a personnel roster. Rows represent workers,
    columns represent days, and each matrix entry \(q_{n,d}\in\{0,1\}\) specifies whether worker
    \(n\) is assigned on day \(d\). Shaded cells correspond to \(q_{n,d}=1\).}
    \label{fig:worker_day_encoding}
\end{figure}
We formulate the personnel-rostering problem as a constrained binary optimization problem and subsequently map it onto the standard QAOA framework. The formulation introduced in this subsection follows the conventional penalty-based construction and serves as the mathematical starting point for the constraint-preserving mixer developed later. We consider a binary personnel-rostering problem with $N$ workers over a planning horizon of $D$ days. The assignment of worker $n$ on day $d$ is represented by the binary decision variable
\begin{equation}
q_{n,d}\in\{0,1\},
\qquad
n=1,\ldots,N,
\qquad
d=1,\ldots,D,
\end{equation}
where $q_{n,d}=1$ indicates that worker $n$ is assigned on day $d$, and $q_{n,d}=0$ otherwise. A complete roster is therefore specified by the binary matrix $q=\{q_{n,d}\}$.
Figure~\ref{fig:worker_day_encoding} illustrates this worker--day binary encoding for a small
example with \(N=4\) workers and \(D=4\) days. Each row corresponds to a worker and each column
corresponds to a day. Shaded entries indicate assigned shifts, i.e., \(q_{n,d}=1\), whereas
unshaded entries indicate \(q_{n,d}=0\). This matrix representation provides the natural
computational basis used in the QAOA formulation, where each binary variable \(q_{n,d}\) is
mapped to one qubit. Each assignment carries an associated cost $c_{n,d}$. The total assignment cost is
\begin{equation}
C(q)
=
\sum_{n=1}^{N}
\sum_{d=1}^{D}
c_{n,d}q_{n,d}.
\label{eq:assignment_cost}
\end{equation}

The schedule is required to satisfy two hard constraints \cite{Hadfield2017HardSoftConstraints}. The first is the daily coverage constraint,
\begin{equation}
\sum_{n=1}^{N}q_{n,d}
=
W_d,
\qquad
d=1,\ldots,D,
\label{eq:coverage_constraint}
\end{equation}
where $W_d$ denotes the workforce required on day $d$. The second is the no-consecutive-duty constraint,
\begin{equation}
q_{n,d}q_{n,d+1}
=
0,
\qquad
n=1,\ldots,N,
\qquad
d=1,\ldots,D-1,
\label{eq:no_consecutive_constraint}
\end{equation}
which prohibits assigning the same worker on two consecutive days. The resulting constrained optimization problem is, therefore,
\begin{align}
\min_q \quad
&
C(q),
\nonumber\\
\text{subject to}\quad
&
\sum_{n=1}^{N}q_{n,d}=W_d,
\qquad
d=1,\ldots,D,
\nonumber\\
&
q_{n,d}q_{n,d+1}=0,
\qquad
n=1,\ldots,N,
\qquad
d=1,\ldots,D-1.
\label{eq:rostering_optimization_problem}
\end{align}
The objective function determines the energetic ordering of the schedules, while the hard constraints determine the admissible search space. In a conventional penalty-based QAOA formulation these two ingredients are combined in the phase Hamiltonian: the objective contributes the physical cost, and the constraints contribute energetic penalties. The construction developed below follows a different route by moving progressively more of the constraint structure into the mixer Hamiltonian itself.
For subsequent construction, it is convenient to distinguish different feasible sectors of the configuration space. The full binary search space is
\begin{equation}
\mathcal H
=
\{0,1\}^{ND}.
\label{eq:full_binary_space}
\end{equation}
The subset satisfying only the coverage constraint is
\begin{equation}
F_{\mathrm{cov}}
=
\left\{
q\in\mathcal H:
\sum_{n=1}^{N}q_{n,d}=W_d,
\quad
d=1,\ldots,D
\right\},
\label{eq:coverage_feasible_sector}
\end{equation}
while the subset satisfying both hard constraints is
\begin{equation}
F_{\mathrm{full}}
=
\left\{
q\in F_{\mathrm{cov}}:
q_{n,d}q_{n,d+1}=0,
\quad
n=1,\ldots,N,
\quad
d=1,\ldots,D-1
\right\}.
\label{eq:fully_feasible_sector}
\end{equation}
These spaces satisfy the nested relation
\(\mathcal H\supset F_{\mathrm{cov}}\supset F_{\mathrm{full}}\),
which provides the geometric framework underlying the mixer constructions developed later. In this hierarchy, progressively stronger notions of feasibility correspond to progressively smaller dynamically accessible subspaces.
The optimization problem is encoded into qubits through the standard correspondence
\[
\hat q_{n,d}
=
\frac{1-Z_{n,d}}{2},
\]
where $Z_{n,d}$ denotes the Pauli-$Z$ operator acting on qubit $(n,d)$ \cite{Lucas2014IsingNP, Glover2019QUBO4OR}. The corresponding objective Hamiltonian is
\begin{equation}
\hat H_{\mathrm{obj}}
=
\sum_{n=1}^{N}
\sum_{d=1}^{D}
c_{n,d}\hat q_{n,d}.
\label{eq:objective_hamiltonian}
\end{equation}
In the conventional penalty formulation, the hard constraints are enforced energetically by augmenting the objective Hamiltonian with penalty operators. The coverage penalty is
\begin{equation}
\hat P_{\mathrm{cov}}
=
\sum_{d=1}^{D}
\left(
\sum_{n=1}^{N}\hat q_{n,d}
-
W_d
\right)^2,
\label{eq:coverage_penalty}
\end{equation}
while the no-consecutive-duty penalty is
\begin{equation}
\hat P_{\mathrm{con}}
=
\sum_{n=1}^{N}
\sum_{d=1}^{D-1}
\hat q_{n,d}\hat q_{n,d+1}.
\label{eq:consecutive_penalty}
\end{equation}
The corresponding QAOA phase Hamiltonian is
\begin{equation}
\hat H_C
=
\hat H_{\mathrm{obj}}
+
A\hat P_{\mathrm{cov}}
+
B\hat P_{\mathrm{con}},
\label{eq:phase_hamiltonian}
\end{equation}
where $A$ and $B$ are positive penalty coefficients.
For a mixer Hamiltonian $\hat H_M$, the QAOA evolution operators are
$U_C(\gamma)
=
e^{-i\gamma\hat H_C},
\qquad
U_M(\beta)
=
e^{-i\beta\hat H_M}$.
The depth-$p$ variational state is
\begin{equation}
|\psi_p(\boldsymbol{\gamma},\boldsymbol{\beta})\rangle
=
\prod_{\ell=1}^{p}
U_M(\beta_\ell)
U_C(\gamma_\ell)
|\psi_0\rangle,
\label{eq:qaoa_state}
\end{equation}
where
$\boldsymbol{\gamma}
=
(\gamma_1,\ldots,\gamma_p), 
\boldsymbol{\beta}
=
(\beta_1,\ldots,\beta_p),$
are optimized by a classical outer loop \cite{Glover2019QUBO4OR, Farhi2014QAOABoundedCSP}.
Up to this point the development follows the standard penalty-based formulation of QAOA. The central question addressed in the remainder of this work is not the variational structure itself, but the construction of the mixer Hamiltonian. Rather than enforcing constraints solely through energetic penalties, we shall embed them directly into the quantum dynamics by constructing mixers whose transition graphs are confined successively to the sectors
\(
\mathcal H, 
F_{\mathrm{cov}}, 
F_{\mathrm{full}}.
\)
This dynamical viewpoint forms the basis of the constraint-preserving framework developed in the following subsections.

\subsection{Constraint-Preserving Dynamics and the Coverage-Preserving Mixer}

The preceding subsection formulated the rostering problem within the standard QAOA framework. The essential degree of freedom that remains is the choice of the mixer Hamiltonian. Although the variational structure of QAOA is independent of the optimization problem, the mixer determines how probability amplitude propagates through the computational basis and therefore determines which schedules are dynamically accessible during the optimization. This observation admits a natural graph-theoretic interpretation. 
Let \(\mathcal B=\{|q\rangle:q\in\mathcal H\}\) denote the computational basis. For a mixer Hamiltonian \(\hat H_M\), we define the transition graph \(\mathcal G(\hat H_M)=(\mathcal B,\mathcal E)\), where an edge \((|q\rangle,|q'\rangle)\in\mathcal E\) is present whenever
\[
\langle q'|\hat H_M|q\rangle\neq 0 .
\]
Since the mixer evolution is generated by \(U_M(\beta)=\exp(-i\beta\hat H_M)\), amplitude can propagate only along the edges of \(\mathcal G(\hat H_M)\). Consequently, the connected component containing the initial state defines the dynamically accessible search space explored by the QAOA circuit. The conventional choice is the transverse-field mixer
\begin{equation}
\hat H_M^{X}
=
\sum_{n=1}^{N}
\sum_{d=1}^{D}
X_{n,d},
\end{equation}
which generates independent single-qubit flips. Its transition graph spans the entire binary configuration space, its vertex set corresponds to \(\mathcal H\)
so neither the daily coverage constraint nor the no-consecutive-duty constraint is preserved during the evolution. Feasibility is therefore enforced solely through the penalty operators appearing in the phase Hamiltonian.

A different strategy is to incorporate constraints directly into the transition operator itself. For the personnel-rostering problem, the most natural starting point is the daily coverage constraint. Since the occupation number
\[
N_d
=
\sum_{n=1}^{N}
q_{n,d}
\]
must remain fixed on every day, admissible elementary transitions should conserve $N_d$.
The simplest such transition is the exchange
\(
|10\rangle
\longleftrightarrow
|01\rangle,
\)
between two workers assigned on the same day, while leaving the configurations
\(
|00\rangle, 
|11\rangle
\)
unchanged. This exchange is generated by the XY interaction
\[
\hat A_{u,v,d}
=
X_{u,d}X_{v,d}
+
Y_{u,d}Y_{v,d},
\]
acting on workers $u$ and $v$ on day $d$ \cite{Hadfield2019QAOAtoQAOAplus, Cook2020QAOAMaxKVertexCover}.

Summing over all connected worker pairs yields the coverage-preserving mixer
\begin{equation}
\hat H_M^{\mathrm{cov}}
=
\sum_{d=1}^{D}
\sum_{(u,v)\in E_d}
\left(
X_{u,d}X_{v,d}
+
Y_{u,d}Y_{v,d}
\right),
\label{eq:coverage_mixer}
\end{equation}
where $E_d$ denotes the worker-connectivity graph for day $d$. The worker-pair connectivity is specified separately for each ansatz. For the reported Guarded-XY results, the local guarded exchanges are applied on a ring worker-pair graph, with the tight-pattern component included where applicable. \cite{Hadfield2021AnalyticalFrameworkQAOA}.

The coverage-preserving property follows immediately from
\(
\left[
\hat N_d,
\hat H_M^{\mathrm{cov}}
\right]
=
0,
\)
where
\(
\hat N_d
=
\sum_{n=1}^{N}
\hat q_{n,d}.
\)
Hence, if the initial state satisfies
\(
\hat N_d
|\psi(0)\rangle
=
W_d
|\psi(0)\rangle,
\)
then
\(
\hat N_d
|\psi(t)\rangle
=
W_d
|\psi(t)\rangle
\)
for all times generated by the mixer. The entire evolution therefore remains confined to the coverage-feasible sector,
\(
|\psi(t)\rangle
\in
F_{\mathrm{cov}}.
\)

\subsection{Guarded Transition Operator}
\label{subsec:guarded_transition_operator}

The Coverage-XY mixer preserves the daily coverage constraint exactly, but it does not guarantee complete schedule feasibility. Although every elementary XY exchange conserves the occupation number on each day, it may still assign the same worker on two consecutive days. Consequently, the transition graph generated by
\(
\hat H_M^{\mathrm{cov}}
\)
is confined to
\(
F_{\mathrm{cov}}
\),
but still contains edges leaving the fully feasible sector
\(
F_{\mathrm{full}}.
\)

The central contribution of this work is the construction of a guarded transition operator that removes these remaining infeasible transitions directly at the level of the mixer Hamiltonian. Instead of proposing every elementary exchange and suppressing infeasible schedules energetically through penalty terms, we modify the transition operator itself so that only exchanges preserving all hard constraints remain dynamically accessible. Feasibility therefore becomes a property of the quantum dynamics rather than a consequence of the optimization landscape.

For an exchange between workers \(u\) and \(v\) on day \(d\), only assignments on the neighboring days can create a new violation of the no-consecutive-duty constraint. We therefore define the neighboring-day set
\[
\mathcal A(d)
=
\{d-1,d+1\}
\cap
\{1,\ldots,D\}.
\]

The guarded mixer used in the numerical simulations employs a symmetric local guard projector,
\begin{equation}
\hat G_{u,v,d}
=
\prod_{\tau\in\mathcal A(d)}
(1-\hat q_{u,\tau})
(1-\hat q_{v,\tau}),
\label{eq:guard_projector}
\end{equation}
which allows an exchange between workers \(u\) and \(v\) on day \(d\) only when both workers are unassigned on the neighboring days. This is a sufficient feasibility-preserving guard: it may remove some otherwise feasible local exchanges, but every exchange that remains preserves the no-consecutive-duty constraint when acting on a fully feasible schedule.

Introducing the spin-exchange operators
\[
\sigma^\pm_{n,d}
=
\frac{X_{n,d}\pm iY_{n,d}}{2},
\]
the guarded mixer is defined as
\begin{equation}
\hat H_M^{\mathrm{guard}}
=
\sum_{d=1}^{D}
\sum_{u<v}
\hat G_{u,v,d}
\left(
\sigma^+_{u,d}\sigma^-_{v,d}
+
\sigma^-_{u,d}\sigma^+_{v,d}
\right).
\label{eq:guarded_mixer}
\end{equation}

Equation~\eqref{eq:guarded_mixer} differs from the conventional Coverage-XY mixer only through the insertion of the local guard projector. The XY exchange continues to enforce conservation of the daily workforce, while the guard removes every elementary transition that would violate the no-consecutive-duty constraint. Consequently, for any \(q\in F_{\mathrm{full}}\),
\[
\langle q'|
\hat H_M^{\mathrm{guard}}
|q\rangle
\neq
0
\quad
\Longrightarrow
\quad
q'
\in
F_{\mathrm{full}},
\]
so every non-zero transition generated from a fully feasible schedule remains within the fully feasible sector.

Therefore, if the initial state satisfies
\[
|\psi(0)\rangle
\in
F_{\mathrm{full}},
\]
then the guarded-mixer evolution satisfies
\[
e^{-it\hat H_M^{\mathrm{guard}}}
|\psi(0)\rangle
\in
F_{\mathrm{full}},
\qquad
\forall t,
\]
and the fully feasible sector becomes an invariant subspace of the quantum dynamics.

An immediate consequence is that the hard scheduling constraints no longer need to appear in the phase Hamiltonian. Since infeasible configurations are never generated during the variational evolution, the phase operator reduces simply to
\[
\hat H_C
=
\hat H_{\mathrm{obj}},
\]
eliminating the need to calibrate hard-constraint penalty coefficients.

The guarded mixer establishes exact feasibility preservation, but local one-day guarded exchanges may be insufficient in saturated workload patterns. This occurs when neighboring days become simultaneously saturated,
\[
W_d+W_{d+1}=N,
\]
forcing complementary worker assignments across adjacent days. In such situations, changing between feasible alternating patterns may require a collective transition across the saturated segment rather than a sequence of local guarded exchanges.

To include these missing feasible transitions while maintaining exact feasibility, we introduce an additional collective transition operator acting on maximal saturated segments. Let
\(
\mathcal T
\)
denote the set of maximal saturated segments. For each segment
\(
C\in\mathcal T,
\)
we define a collective alternating-pattern exchange operator
\(
\hat A_{u,v}^{C}
\)
together with a corresponding boundary guard
\(
\hat G_{u,v}^{C}.
\)
The resulting tight-pattern contribution is
\begin{equation}
\hat H_M^{\mathrm{tight}}
=
\sum_{C\in\mathcal T}
\sum_{u<v}
\hat G^{C}_{u,v}
\left(
\hat A^{C}_{u,v}
+
\hat A^{C\dagger}_{u,v}
\right),
\label{eq:tight_mixer}
\end{equation}
whose operator-level construction and feasibility rationale are given in Appendix~\ref{app:tight_pattern_mixer}.

The complete constraint-preserving mixer introduced in this work is therefore
\begin{equation}
\hat H_M
=
\hat H_M^{\mathrm{guard}}
+
\hat H_M^{\mathrm{tight}}.
\label{eq:complete_mixer}
\end{equation}

The first term guarantees exact preservation of both hard scheduling constraints throughout the variational evolution, while the second adds collective feasible exchanges in saturated workload segments where local guarded exchanges alone are insufficient. Together they generate a transition graph that is confined to the fully feasible sector while retaining additional feasible transitions needed for exploration. The resulting framework separates two complementary design principles for constrained QAOA---\emph{feasibility preservation} and \emph{feasible-transition design}. The guarded mixer enforces feasibility by eliminating infeasible transitions, while the tight-pattern operator supplements the local guarded exchanges with collective feasible moves required by saturated workload patterns.

\section{Benchmark and Simulation Protocol}

The mathematical construction developed in the previous section defines three levels of variational dynamics: unconstrained evolution on the full binary Hilbert space, coverage-preserving evolution on the coverage-feasible subspace, and fully constraint-preserving evolution on the fully feasible subspace. The purpose of the numerical study is to compare these three approaches under identical benchmark instances, optimization protocols, and performance diagnostics.

We consider three QAOA ansätze. The Penalty-X ansatz employs the conventional transverse-field mixer and explores the full binary Hilbert space $\mathcal H$, with both hard constraints enforced through energetic penalty terms in the phase Hamiltonian \cite{Farhi2014QAOABoundedCSP,Farhi2014QAOA}. The Coverage-XY ansatz employs the excitation-conserving XY mixer and, when initialized in $F_{\mathrm{cov}}$, preserves the daily coverage constraint throughout the variational evolution so that the dynamics remain confined to $F_{\mathrm{cov}}$ \cite{Hadfield2019QAOAtoQAOAplus}. However, it still requires an energetic penalty to suppress violations of the no-consecutive-duty constraint. The proposed Guarded-XY ansatz employs the guarded transition operator developed in Sec.~\ref{subsec:guarded_transition_operator}. When initialized in $F_{\mathrm{full}}$, its evolution remains entirely within the fully feasible subspace, eliminating the need for hard-constraint penalty terms \cite{Wang2020XYMixers}.

Thus, the numerical comparison is not merely between different mixer Hamiltonians; it is a comparison between different geometries of variational exploration,
\begin{equation}
\mathcal H
\supset
F_{\mathrm{cov}}
\supset
F_{\mathrm{full}},
\end{equation}
where $\mathcal H$ denotes the full binary configuration space, $F_{\mathrm{cov}}$ the coverage-feasible sector, and $F_{\mathrm{full}}$ the fully feasible sector satisfying both hard scheduling constraints.

\subsection{Benchmark Instance and Feasible-Space Enumeration}
We use \(N=4\) workers and \(D=4\) days with workload \(W=[2,1,2,1]\). The full binary space contains \(2^{ND}=2^{16}=65536\) schedules. Exact enumeration gives \(|F_{\mathrm{cov}}|=576\) and \(|F_{\mathrm{full}}|=72\), so only \(|F_{\mathrm{full}}|/|F_{\mathrm{cov}}|=0.125\) of coverage-feasible schedules are fully feasible. This reduction makes the instance useful for distinguishing partial constraint preservation from full constraint preservation: a coverage-preserving mixer still explores many schedules that violate the no-consecutive-duty constraint, whereas a guarded mixer is restricted to the smaller fully feasible roster subspace.

Since the benchmark is small, the fully feasible space can be enumerated exactly. This gives
\[
C_{\min}=5.011797,
\qquad
C_{\mathrm{mean}}=7.369879,
\qquad
C_{\max}=9.379095.
\]
The optimal schedule is
\(
0011|0100|1010|0001,
\)
where the vertical bars separate the four scheduling days. This schedule is not supplied to the optimizer and is used only for post-processing diagnostics.

A preliminary depth-one Coverage-XY run gives an output-averaged normalized score
\(
\langle r\rangle
\approx
0.303,
\)
showing that a single shallow layer does not already solve the instance. The selected benchmark is therefore small enough for exact statevector simulation and exhaustive diagnostics, while still separating the behaviour of penalty-based, partially constraint-preserving, and fully constraint-preserving QAOA dynamics.

\subsection{Penalty-Coefficient Calibration}

The three ansatz variants considered in this work differ not only in their mixer Hamiltonians but also in the role played by energetic penalty terms. The Penalty-X ansatz enforces both hard constraints through the phase Hamiltonian, the Coverage-XY ansatz requires a penalty only for the no-consecutive-duty constraint, whereas the Guarded-XY ansatz preserves both hard constraints dynamically and therefore requires no hard-constraint penalties. A consistent comparison therefore requires calibrating the penalty coefficients only for those ans\"atze in which they are present.

The purpose of the penalty calibration is to prevent infeasible schedules with artificially low assignment cost from competing energetically with high-quality feasible schedules. At the same time, the penalties should not be chosen unnecessarily large, since excessive penalties dominate the assignment-cost landscape and reduce the effective resolution of the variational optimization.

We define the reference threshold
\[
T
=
\frac{C_{\min}+C_{\max}}{2}
=
7.195446.
\]
This threshold lies between the optimal feasible cost and the upper part of the feasible spectrum and serves as a practical boundary separating low-cost feasible schedules from the remainder of the feasible subspace. The penalty strengths are then chosen so that infeasible schedules with dangerously low assignment cost are shifted above this threshold.

For the Coverage-XY ansatz, the daily coverage constraint is already preserved by the mixer. Consequently, only the no-consecutive-duty constraint must be enforced energetically, and the phase Hamiltonian is
\begin{equation}
H_{\mathrm{covXY}}
=
C
+
B P_{\mathrm{con}}.
\end{equation}
We enumerate all schedules belonging to $F_{\mathrm{cov}}$ that violate the no-consecutive-duty constraint and determine the smallest value of $B$ that raises the low-cost violating schedules above the threshold $T$. This procedure gives
\(
B
=
2.460466.
\)

For the Penalty-X baseline, neither hard constraint is preserved dynamically, so both must be incorporated into the phase Hamiltonian,
\begin{equation}
H_{\mathrm{penX}}
=
C
+
A P_{\mathrm{cov}}
+
B P_{\mathrm{con}}.
\end{equation}
Applying the same calibration procedure gives
\(
A
=
2.703661, 
B
=
2.460466.
\)

For the Guarded-XY ansatz, both hard constraints are enforced directly by the guarded mixer. Consequently, no hard-constraint penalties are required and the phase Hamiltonian reduces simply to
$H_{\mathrm{guard}}
=
C.$

This distinction illustrates the different philosophies underlying the three approaches. Penalty-X relies entirely on energetic suppression of infeasible schedules. Coverage-XY combines dynamical preservation of daily staffing with energetic suppression of no-consecutive-duty violations. Guarded-XY removes infeasible schedules from the dynamics altogether, thereby eliminating the need to tune hard-constraint penalty coefficients.

As an independent consistency check, the Coverage-XY penalty coefficient was also estimated using a CP-SAT formulation that minimizes the assignment cost over coverage-feasible schedules at fixed numbers of consecutive-duty violations. This independent optimization gives $B
\approx
2.4605,$ in excellent agreement with the exact enumeration result. For larger benchmark instances, where exhaustive enumeration is impractical, the same CP-SAT procedure can therefore be employed as a time-limited calibration method.

\subsection{Hamiltonian Rescaling and Parameter Initialization}

After fixing the phase Hamiltonians for the three ans\"atze, we calibrate an overall Hamiltonian rescaling factor for each construction. The phase-separation operator is written as
\begin{equation}
U_C(\gamma_\ell)
=
\exp
\!\left[
-i\gamma_\ell
\lambda_s
H_C
\right],
\end{equation}
where $H_C$ denotes the ansatz-dependent phase Hamiltonian and $\lambda_s$ is a scalar rescaling factor.

The purpose of $\lambda_s$ is not to modify the optimization problem or alter the classical optimum. Instead, it establishes a comparable phase scale for the variational parameter $\gamma_\ell$. Since the Penalty-X, Coverage-XY, and Guarded-XY ans\"atze employ phase Hamiltonians with substantially different energy scales, using a common phase normalization would produce variational landscapes with very different resolutions \cite{sureshbabu2024parametersetting}. Penalty-X contains two hard-constraint penalties, Coverage-XY contains only one, while Guarded-XY contains none. Consequently, the phase accumulated during the evolution differs significantly among the three constructions.

To calibrate the rescaling factor, we perform depth-one scans of the assignment-cost expectation value
\begin{equation}
\langle C\rangle_{\gamma,\beta}
=
\langle
\psi_1(\gamma,\beta)
|
C
|
\psi_1(\gamma,\beta)
\rangle,
\end{equation}
where
\[
|\psi_1(\gamma,\beta)\rangle
=
U_M(\beta)
\exp
\!\left[
-i\gamma
\lambda_s
H_C
\right]
|\psi_0\rangle .
\]

\begin{figure*}[t]
    \centering
    \begin{subfigure}[t]{0.48\textwidth}
        \centering
        \includegraphics[width=\linewidth]
        {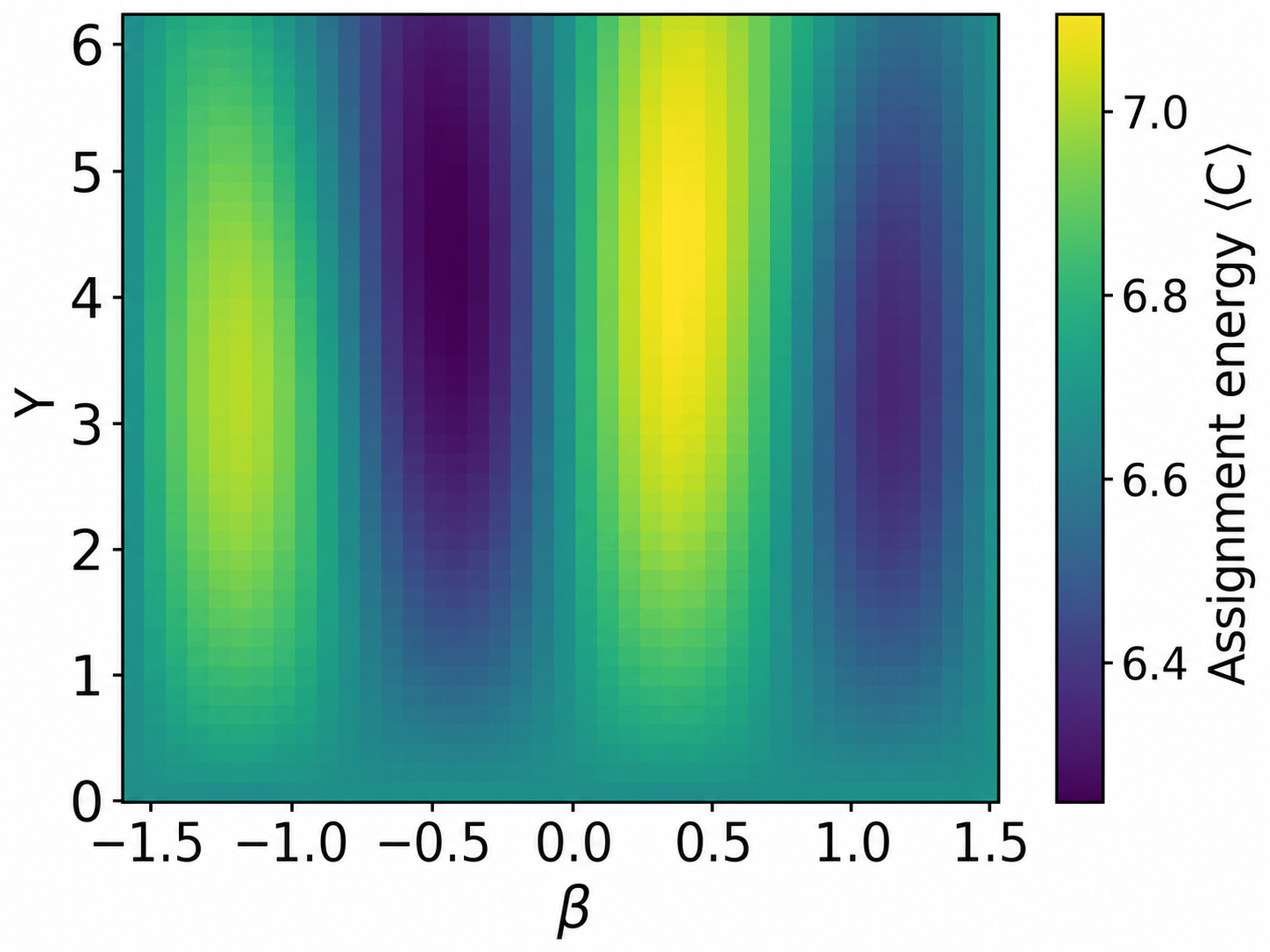}
        \caption{Coverage-XY.}
        \label{fig:coverage_xy_lambda_heatmap}
    \end{subfigure}
    \hfill
    \begin{subfigure}[t]{0.48\textwidth}
        \centering
        \includegraphics[width=\linewidth]
        {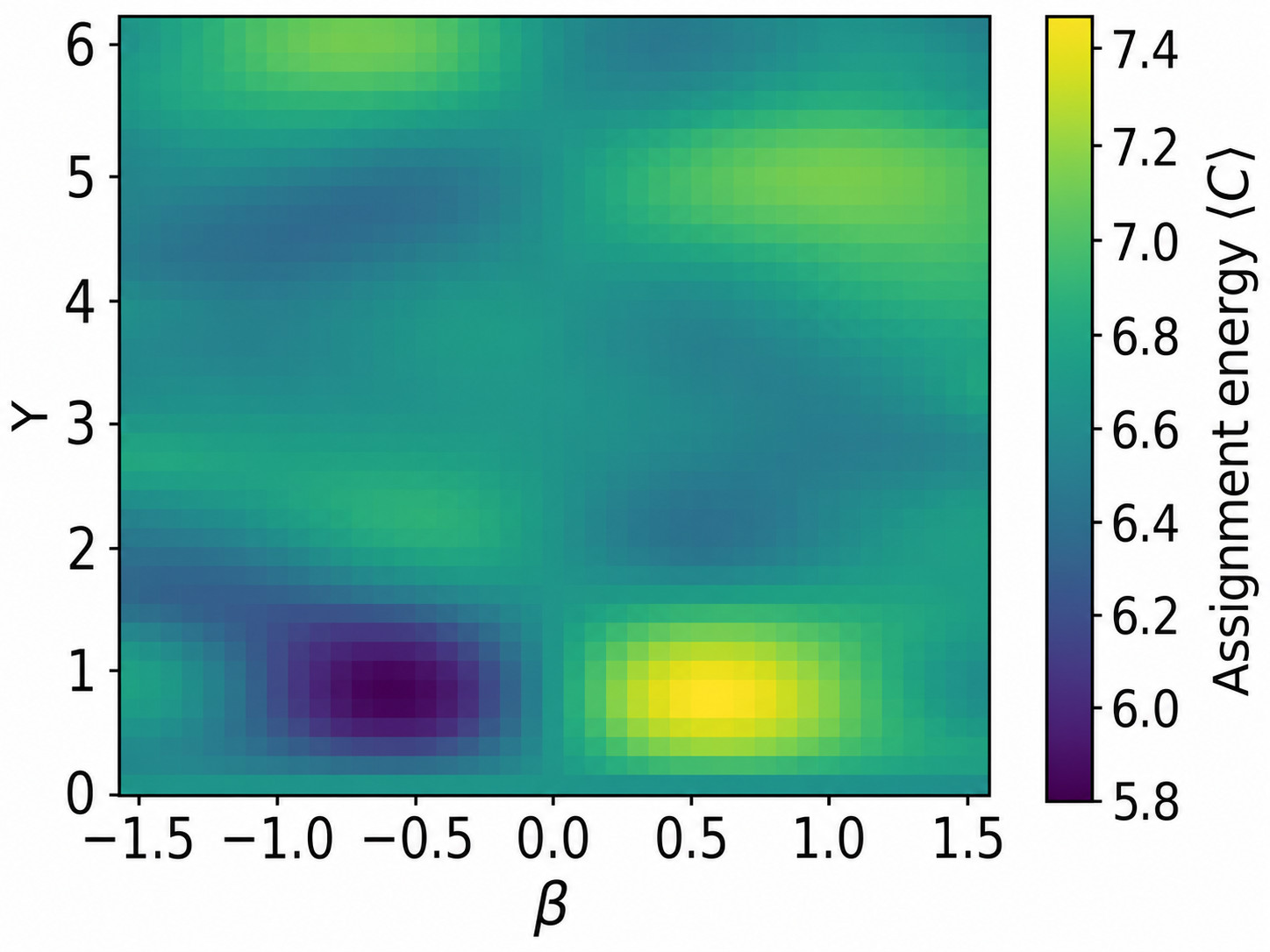}
        \caption{Guarded-XY.}
        \label{fig:guarded_xy_lambda_heatmap}
    \end{subfigure}
    \caption{
    Depth-one assignment-cost landscapes used to calibrate the Hamiltonian rescaling factor. The selected values are $\lambda_s=0.1$ for Coverage-XY and $\lambda_s=2.25$ for Guarded-XY.}
    \label{fig:lambda_landscapes}
\end{figure*}

The assignment-cost expectation is used rather than the training Hamiltonian so that all ans\"atze are evaluated using the same physical objective function. The resulting depth-one landscapes are shown in Fig.~\ref{fig:lambda_landscapes}.
The chosen value of $\lambda_s$ should produce a landscape that is sufficiently structured for optimization while avoiding excessive oscillations. If $\lambda_s$ is too small, the phase separator weakly perturbs the initial state and the landscape becomes nearly flat. Conversely, if $\lambda_s$ is too large, rapid phase oscillations generate numerous shallow local extrema that complicate classical optimization.

For the Coverage-XY ansatz, the remaining no-consecutive-duty penalty already increases the overall energy scale, so a relatively small rescaling factor, $\lambda_s^{\mathrm{covXY}}
=
0.1,$ produces a well-resolved low-cost basin. In contrast, the Guarded-XY phase Hamiltonian contains only the assignment cost restricted to the fully feasible subspace. A substantially larger value, $\lambda_s^{\mathrm{guard}}
=
2.25,$ 
is therefore required to obtain comparable phase variation.

The same depth-one landscapes are also used to initialize the classical optimization. Candidate starting points are selected from the low-cost regions of the $(\gamma,\beta)$ plane and supplemented by random initializations. For circuit depths $p>1$, optimized parameters obtained at depth $p-1$ are extended by one additional layer and used as warm-start initial guesses \cite{sack2021tqa}. Thus, Hamiltonian rescaling and parameter initialization play complementary roles, the former produces a numerically well-resolved variational landscape, while the latter improves the robustness and reproducibility of the classical optimization.

\subsection{Initial States and Simulation Configurations}

The initial state associated with each ansatz is chosen to respect the constraints preserved by its mixer Hamiltonian. Since the mixer can only explore the connected component containing the initial state, this choice ensures that the accessible search space is consistent with the intended variational dynamics \cite{Hadfield2019QAOAtoQAOAplus, Hadfield2021AnalyticalFrameworkQAOA}.

For the Penalty-X baseline, the initial state is the uniform superposition over the full binary Hilbert space,
\begin{equation}
|\psi_0^{X}\rangle
=
|+\rangle^{\otimes ND}.
\end{equation}

For the Coverage-XY ansatz, we employ the uniform superposition over the coverage-feasible sector,
\begin{equation}
|\psi_0^{\mathrm{cov}}\rangle
=
|F_{\mathrm{cov}}|^{-1/2}
\sum_{q\in F_{\mathrm{cov}}}
|q\rangle .
\end{equation}

For the Guarded-XY ansatz, the initial state is the uniform superposition over the fully feasible sector,
\begin{equation}
|\psi_0^{\mathrm{guard}}\rangle
=
|F_{\mathrm{full}}|^{-1/2}
\sum_{q\in F_{\mathrm{full}}}
|q\rangle .
\end{equation}

To investigate initialization effects, we additionally consider a localized feasible starting schedule, $q_{\mathrm{init}}
=
0101|0010|1001|0100,$
whose assignment cost is $C(q_{\mathrm{init}})
=
8.881886.$

Before the variational circuit, a small number of mixer-only spreading layers are applied,
\begin{equation}
|\psi_0^{\mathrm{loc}}(L_{\mathrm{init}})\rangle
=
\left[
U_M^{\mathrm{GXY}}
(\beta_{\mathrm{init}})
\right]^{L_{\mathrm{init}}}
|q_{\mathrm{init}}\rangle ,
\end{equation}
with $\beta_{\mathrm{init}}
=
0.2, L_{\mathrm{init}}
\in
\{1,2,3\}.$
Throughout this work, all structured initial states are prepared directly as normalized statevectors. This allows the numerical comparison to isolate the effects of mixer design and feasible-subspace dynamics without introducing additional circuit overhead associated with state preparation. Coverage-XY employs complete same-day connectivity, while for the reported Guarded-XY results, local guarded exchanges are applied on a ring worker-pair graph, with the tight-pattern component included where applicable. For the benchmark workload $W=[2,1,2,1],$ no saturated adjacent-day segments occur, so the combined Guarded-XY+tight mixer reduces to the guarded one-day exchange construction developed in the previous section. In the structural sensitivity study of Sec.~\ref{subsec:guarded_structural_sensitivity}, however, workloads with nonzero \(N_{\mathrm{tight}}\) are considered, and the Guarded-XY+tight construction is used to include the corresponding collective tight-pattern exchanges.
The simulation configurations used throughout this work are summarized in Table~\ref{tab:simulation_configurations}.
\begin{table}[t]
\centering
\caption{Initial-state and constraint-handling configurations used in the simulations.}
\label{tab:simulation_configurations}
\scriptsize
\setlength{\tabcolsep}{3pt}
\renewcommand{\arraystretch}{1.18}

\begin{tabularx}{\textwidth}{
@{}
>{\raggedright\arraybackslash}p{0.16\textwidth}
>{\raggedright\arraybackslash}p{0.28\textwidth}
>{\raggedright\arraybackslash}p{0.20\textwidth}
>{\raggedright\arraybackslash}X
@{}}
\toprule
Method & Initial state & Mixer & Constraint handling \\
\midrule

Penalty-X &
\( |+\rangle^{\otimes ND} \) &
Transverse-field \(X\) &
Both hard constraints penalized \\

Coverage-XY &
\(\displaystyle |F_{\mathrm{cov}}|^{-1/2}
\sum_{q\in F_{\mathrm{cov}}}|q\rangle\) &
Same-day XY &
Coverage preserved; no-consecutive-duty penalized \\

Guarded-XY &
\(\displaystyle |F_{\mathrm{full}}|^{-1/2}
\sum_{q\in F_{\mathrm{full}}}|q\rangle\) &
Guarded XY; Guarded-XY+tight\({}^{\dagger}\) &
Both hard constraints preserved \\

Guarded-XY localized &
\(\displaystyle
[U_M^{\mathrm{GXY}}(\beta_{\mathrm{init}})]^{L_{\mathrm{init}}}
|q_{\mathrm{init}}\rangle\) &
Guarded XY &
Both hard constraints preserved \\

\bottomrule
\end{tabularx}

\vspace{0.4em}
\raggedright
\footnotesize
\({}^{\dagger}\) For the main benchmark \(W=[2,1,2,1]\), no saturated adjacent-day segment is present, so Guarded-XY+tight reduces to the one-day Guarded-XY mixer. In the structural sensitivity study of Sec.~\ref{subsec:guarded_structural_sensitivity}, workloads with nonzero \(N_{\mathrm{tight}}\) use the tight-pattern component.
\end{table}

\subsection{Optimization Protocol and Evaluation Metrics}
\label{subsec:optimization_protocol}
For each ansatz, we optimize the depth-\(p\) QAOA state
\begin{equation}
|\psi_p(\boldsymbol{\gamma},\boldsymbol{\beta})\rangle
=
\prod_{\ell=1}^{p}
U_M(\beta_\ell)
U_C(\gamma_\ell)
|\psi_0\rangle,
\end{equation}
for depths \(p=1,\ldots,6\). Two optimization objectives are considered. The first is the standard expectation-value objective, which minimizes the average value of the training Hamiltonian. The second is the Conditional Value-at-Risk (CVaR) objective, which averages only the lowest-energy fraction \(\alpha\) of the measured energy distribution. Throughout this work we use $\alpha=0.2.$
Expectation-value optimization therefore probes the full output distribution, whereas CVaR emphasizes its low-energy tail.

The classical optimization uses a combination of landscape-based initial guesses, random initializations, and parameter continuation from lower depths. For \(p>1\), optimized parameters from depth \(p-1\) are extended by one additional layer and used as candidate starting points. All simulations are performed using exact statevector evolution, so sampling noise, gate errors, and decoherence are not included.

After optimization, the final output distribution is
\begin{equation}
P(q)
=
\left|
\langle q|
\psi_p(\boldsymbol{\gamma}^{*},\boldsymbol{\beta}^{*})
\rangle
\right|^2 ,
\end{equation}
where \(\boldsymbol{\gamma}^{*}\) and \(\boldsymbol{\beta}^{*}\) are the optimized parameters. We evaluate this distribution using feasibility and solution-quality diagnostics. The coverage-feasibility and full-feasibility probabilities are
\begin{equation}
P_{\mathrm{cov}}
=
\sum_{q\in F_{\mathrm{cov}}}P(q),
\qquad
P_{\mathrm{full}}
=
\sum_{q\in F_{\mathrm{full}}}P(q).
\end{equation}
The optimal-state probability is
\begin{equation}
P_{\mathrm{opt}}
=
\sum_{q\in F_{\mathrm{opt}}}P(q),
\end{equation}
where \(F_{\mathrm{opt}}\) is the set of fully feasible schedules with minimum assignment cost \(C_{\min}\). We also use the feasibility-conditioned optimal probability
\begin{equation}
P_{\mathrm{opt}\mid\mathrm{full}}
=
\frac{P_{\mathrm{opt}}}{P_{\mathrm{full}}},
\end{equation}
when \(P_{\mathrm{full}}>0\), and set it to zero otherwise.

To measure average feasible-solution quality, we define the normalized score
\begin{equation}
r(q)
=
\frac{C_{\max}-C(q)}
{C_{\max}-C_{\min}},
\qquad q\in F_{\mathrm{full}},
\end{equation}
where \(C_{\min}\) and \(C_{\max}\) are the minimum and maximum assignment costs over \(F_{\mathrm{full}}\). Thus, \(r(q)=1\) for an optimal feasible schedule and \(r(q)=0\) for the worst feasible schedule. Infeasible schedules are assigned zero score. The output-averaged score is
\begin{equation}
\langle r\rangle
=
\sum_q P(q)r(q).
\end{equation}

Together, \(P_{\mathrm{cov}}\), \(P_{\mathrm{full}}\), \(P_{\mathrm{opt}}\), \(P_{\mathrm{opt}\mid\mathrm{full}}\), and \(\langle r\rangle\) distinguish feasibility preservation, exact-optimum concentration, and average output-distribution quality.
\section{Results and Discussion}

\subsection{Depth Dependence of Optimal-State Concentration}

We first compare the three QAOA variants through the optimal-state probability
\(P_{\mathrm{opt}}\), defined as the probability assigned to the exact optimal feasible roster after variational optimization. This diagnostic directly measures how strongly the output distribution concentrates on the best feasible schedule.

\begin{figure*}[t]
    \centering

    \begin{subfigure}[t]{0.48\textwidth}
        \centering
        \includegraphics[width=\linewidth]
{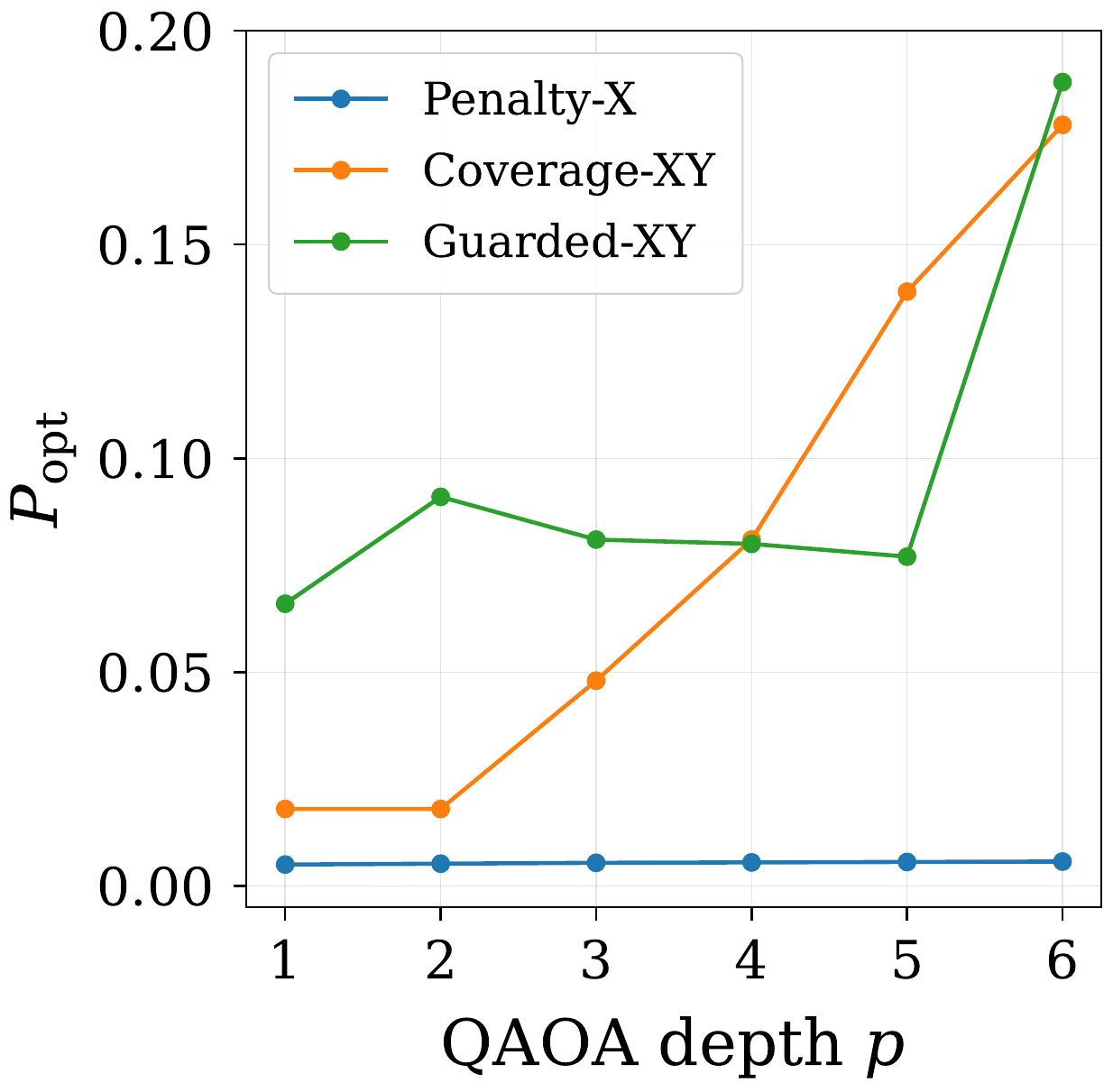}
        \caption{Expectation-value QAOA.}
        \label{fig:expectation_popt_depth}
    \end{subfigure}
    \hfill
    \begin{subfigure}[t]{0.48\textwidth}
        \centering
        \includegraphics[width=\linewidth]
        {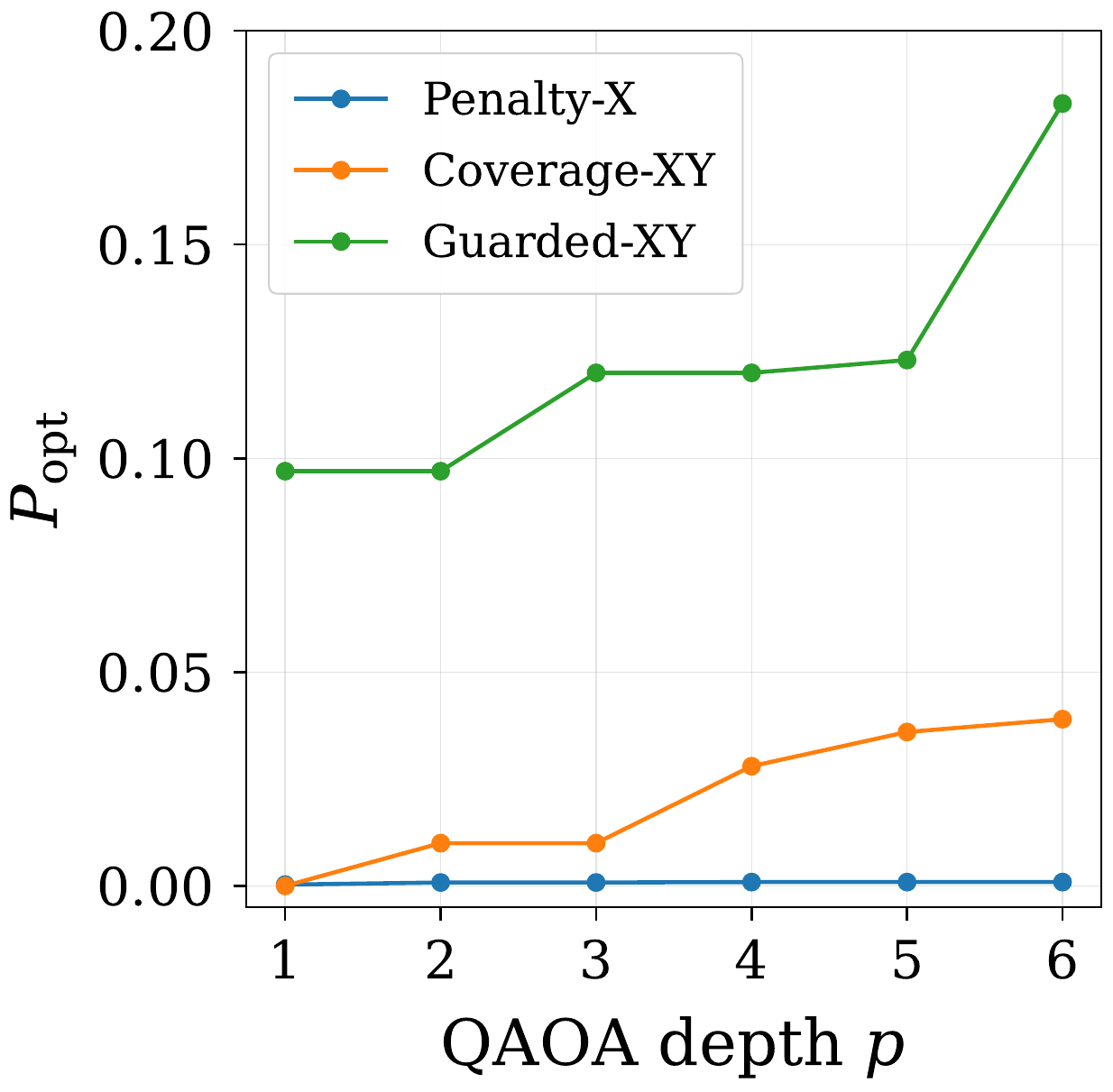}
        \caption{CVaR-QAOA with \(\alpha=0.2\).}
        \label{fig:cvar_popt_depth}
    \end{subfigure}

    \caption{
    Optimal-state probability \(P_{\mathrm{opt}}\) as a function of QAOA depth.
    Panel (a) shows expectation-value optimization and panel (b) shows CVaR optimization.
    }
    \label{fig:popt_depth_sweep}
\end{figure*}

Figure~\ref{fig:popt_depth_sweep} shows \(P_{\mathrm{opt}}\) as a function of QAOA depth for both expectation-value optimization and CVaR optimization ~\cite{Farhi2014QAOA, Barkoutsos2020CVaR}. The comparison uses the calibrated penalties and Hamiltonian rescaling factors described in the previous section. Penalty-X explores the full binary space with both hard constraints penalized. Coverage-XY preserves daily coverage but retains a penalty for no-consecutive-duty violations. Guarded-XY evolves directly within \(F_{\mathrm{full}}\). For the main workload \(
W=[2,1,2,1],
\) there is no saturated adjacent-day segment, so the Guarded-XY+tight construction reduces to the one-day guarded exchange mixer.

In both panels of Fig.~\ref{fig:popt_depth_sweep}, the Penalty-X baseline remains close to zero across the entire depth range. Although the penalty terms energetically suppress infeasible schedules, the transverse-field mixer still spreads probability over the full \(2^{ND}\)-dimensional binary space. As a result, only a small fraction of the output distribution reaches the exact feasible optimum.

The two constraint-preserving ans\"atze perform substantially better. Under expectation-value optimization in Fig.~\ref{fig:popt_depth_sweep}(a), Coverage-XY increases steadily with depth and reaches
\(
P_{\mathrm{opt}}
=
0.184772
\)
at \(p=6\). Guarded-XY is less monotonic at intermediate depths, but reaches
\(
P_{\mathrm{opt}}
=
0.190018
\)
at the same depth. Thus, at the largest depth studied, Guarded-XY is competitive with Coverage-XY in exact-optimum sampling.

This comparison should be interpreted carefully. Coverage-XY is evaluated here under favorable conditions because the no-consecutive-duty penalty coefficient is calibrated exactly for this small enumerable instance. For larger instances, where exhaustive calibration is not available, this coefficient would have to be estimated by a heuristic or time-limited classical procedure. Guarded-XY avoids this sensitivity because both hard constraints are enforced directly by the mixer.

The separation between the two constraint-preserving strategies becomes clearer under CVaR optimization in Fig.~\ref{fig:popt_depth_sweep}(b). With \(\alpha=0.2\), Guarded-XY gives the largest \(P_{\mathrm{opt}}\) at every depth and reaches
\(
P_{\mathrm{opt}}
=
0.189800
\)
at \(p=6\). Coverage-XY also improves with depth, but reaches only
\(
P_{\mathrm{opt}}
=
0.041121,
\)
while Penalty-X remains negligible. Thus, even with exact penalty calibration, the residual penalty-based constraint handling in Coverage-XY is less effective under the tail-focused objective. Guarded-XY keeps the entire variational search inside the fully feasible subspace, so the low-energy tail selected by CVaR is drawn entirely from feasible schedules.

\subsection{Output-Distribution Quality at Maximum Depth}
\begin{figure*}[t]
    \centering
    \begin{subfigure}[t]{0.49\textwidth}
        \centering
        \includegraphics[width=\linewidth]
        {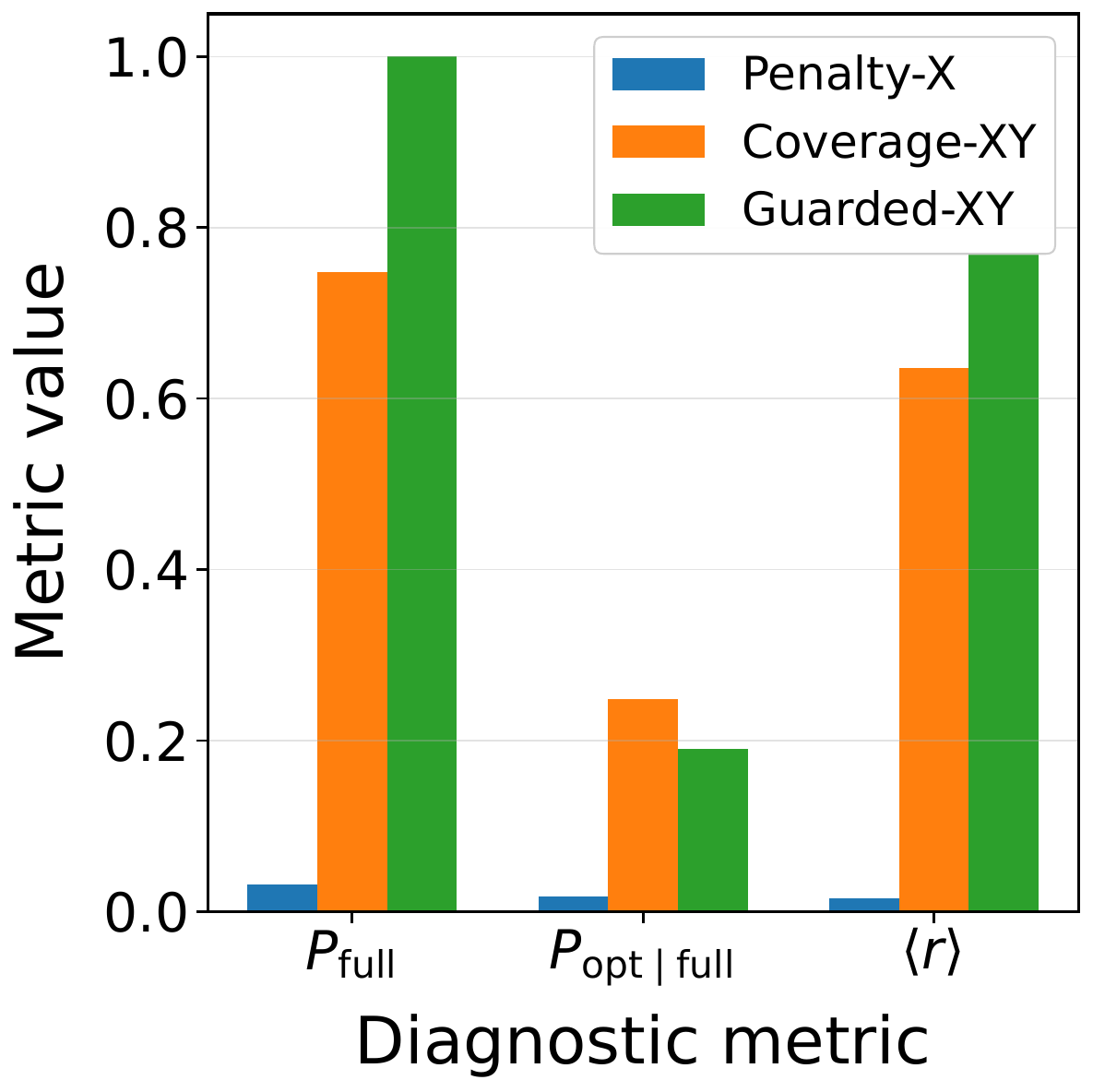}
        \caption{Expectation-value QAOA.}
        \label{fig2a}
    \end{subfigure}
    \hfill
    \begin{subfigure}[t]{0.49\textwidth}
        \centering
        \includegraphics[width=\linewidth]
        {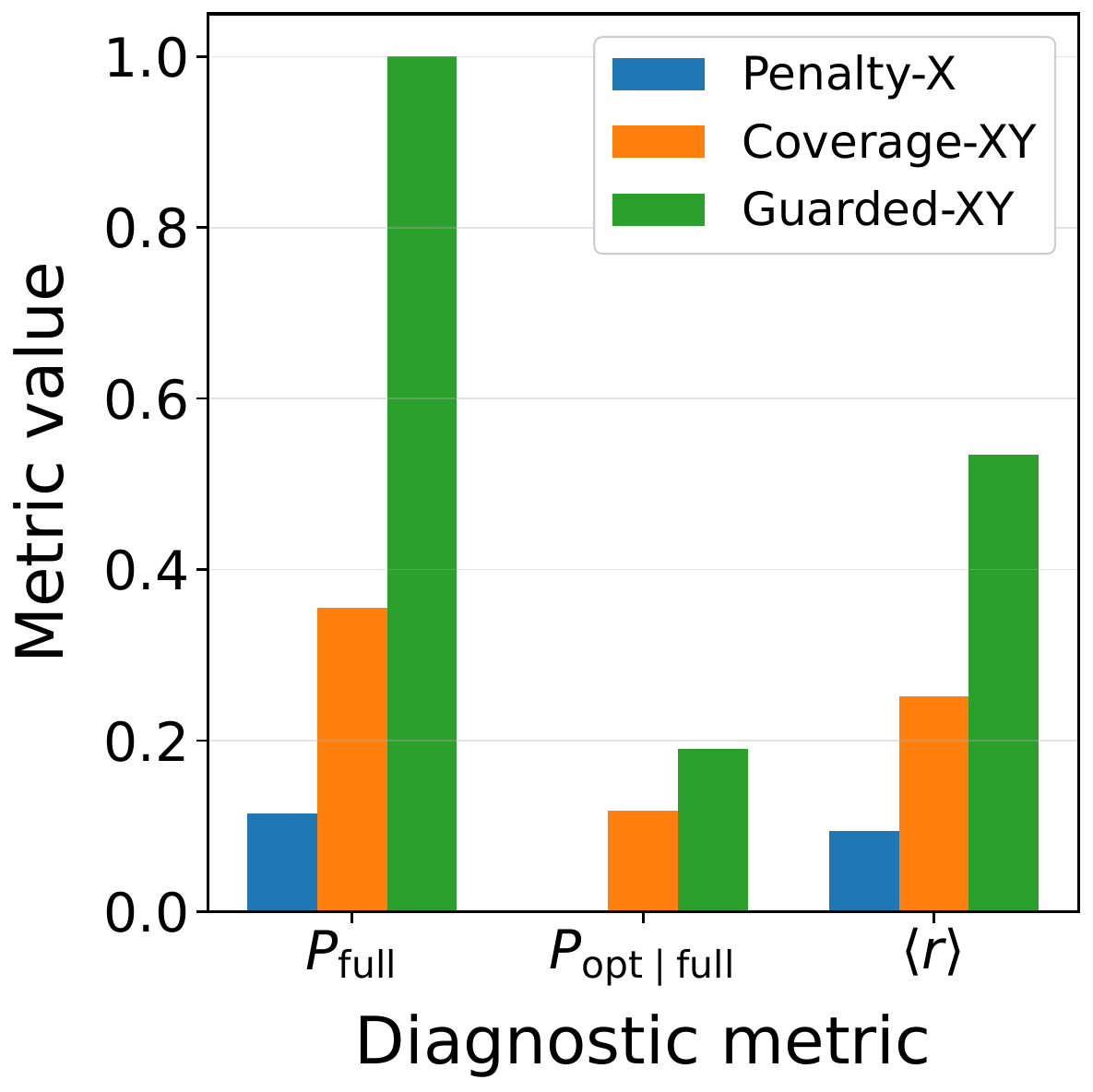}
        \caption{CVaR-QAOA ($\alpha=0.2$).}
        \label{fig:cvar_p6_summary}
    \end{subfigure}
    \caption{
    Distribution-quality diagnostics at the maximum simulated depth \(p=6\). Shown are the full-feasibility probability \(P_{\mathrm{full}}\), the feasibility-conditioned optimal probability \(P_{\mathrm{opt}\mid\mathrm{full}}\), and the output-averaged normalized score \(\langle r\rangle\).
    }
    \label{fig:p6_distribution_summary}
\end{figure*}

The optimal-state probability \(P_{\mathrm{opt}}\) measures concentration on the exact optimum, but it does not fully characterize the quality of the output distribution. A variational state may assign significant probability to the optimal schedule while simultaneously distributing weight over infeasible configurations or lower-quality feasible schedules. We therefore examine the optimized distributions at the maximum simulated depth, \(p=6\), using the diagnostics defined in Sec.~\ref{subsec:optimization_protocol}: \(P_{\mathrm{full}}\), \(P_{\mathrm{opt}\mid\mathrm{full}}\), and \(\langle r\rangle\). The corresponding results are summarized in Fig.~\ref{fig:p6_distribution_summary}.

\begin{table*}[t]
\centering
\caption{
Five most probable computational-basis states at depth \(p=6\) for expectation-value QAOA. Coverage-XY can still assign probability to schedules violating the no-consecutive-duty constraint, whereas Guarded-XY remains entirely within the fully feasible subspace.
}
\label{tab:top_states_coverage_guarded_p6}

\scriptsize
\setlength{\tabcolsep}{3.5pt}
\renewcommand{\arraystretch}{1.15}

\begin{tabular}{ll l c c c c c}
\toprule
Method & Rank & State & Prob. & $C(q)$ & $P_{\mathrm{con}}$ & Feasible & Optimal \\
\midrule

Coverage-XY
& 1 & \texttt{0011|0100|1010|0001} & 0.184772 & 5.011797 & 0 & Yes & Yes \\
& 2 & \texttt{0011|1000|0110|0001} & 0.121458 & 5.334364 & 0 & Yes & No \\
& 3 & \texttt{1001|0100|1010|0001} & 0.095899 & 5.268180 & 0 & Yes & No \\
& 4 & \texttt{0011|0100|0110|0001} & 0.064420 & 3.730372 & 1 & No & No \\
& 5 & \texttt{1001|0010|1100|0001} & 0.058402 & 5.731321 & 0 & Yes & No \\

\midrule

Guarded-XY
& 1 & \texttt{0011|0100|1010|0001} & 0.190018 & 5.011797 & 0 & Yes & Yes \\
& 2 & \texttt{1001|0010|1100|0001} & 0.177503 & 5.731321 & 0 & Yes & No \\
& 3 & \texttt{0011|1000|0110|0001} & 0.143791 & 5.334364 & 0 & Yes & No \\
& 4 & \texttt{0011|0100|1001|0010} & 0.056754 & 7.150008 & 0 & Yes & No \\
& 5 & \texttt{1100|0010|1100|0001} & 0.041126 & 6.612882 & 0 & Yes & No \\

\bottomrule
\end{tabular}
\end{table*}

Under expectation-value optimization, shown in Fig.~\ref{fig:p6_distribution_summary}(a), Coverage-XY exhibits a relatively large conditional concentration on the optimum,
\(
P_{\mathrm{opt}\mid\mathrm{full}}
=
0.246859,
\)
but only
\(
P_{\mathrm{full}}
=
0.748492.
\)
Approximately one quarter of the output distribution therefore lies outside the fully feasible subspace. This reflects the fact that Coverage-XY preserves only the daily coverage constraint. Although all sampled schedules satisfy the staffing requirement, the mixer can still generate schedules violating the no-consecutive-duty constraint.

Guarded-XY exhibits a qualitatively different behaviour. Since its dynamics are confined to \(F_{\mathrm{full}}\) by construction, $P_{\mathrm{full}}
=
1.$
Consequently,
\(
P_{\mathrm{opt}\mid\mathrm{full}}
=
P_{\mathrm{opt}}
=
0.190018.
\)
Although its conditional optimal probability is slightly smaller than that of Coverage-XY, Guarded-XY achieves the largest average solution quality,
\(
\langle r\rangle
=
0.771003.
\)
Rather than concentrating probability on only a few configurations, the guarded dynamics distribute probability over a broad set of high-quality feasible schedules.

The same qualitative behaviour becomes even more pronounced under CVaR optimization, shown in Fig.~\ref{fig:p6_distribution_summary}(b). Coverage-XY yields $P_{\mathrm{full}}
=
0.352850,$ indicating substantial leakage into infeasible schedules. By contrast, Guarded-XY again satisfies $P_{\mathrm{full}}
=
1,$ while simultaneously giving the largest values of both
\(P_{\mathrm{opt}\mid\mathrm{full}}\)
and
\(\langle r\rangle\).
These results demonstrate that once feasibility and solution quality are considered simultaneously, the fully constraint-preserving ansatz produces the highest-quality output distribution among the three approaches. The origin of these differences can be understood more directly by examining the most probable computational-basis states generated by the two constraint-preserving ans\"atze. These are listed in Table~\ref{tab:top_states_coverage_guarded_p6}.
Table~\ref{tab:top_states_coverage_guarded_p6} explains the different behaviour observed in Fig.~\ref{fig:p6_distribution_summary}. Both ans\"atze assign their largest probability to the exact optimal schedule, showing that each is capable of locating the global optimum. The distinction lies not in identifying the optimum itself, but in how the remaining probability is distributed over the accessible search space.

Coverage-XY assigns substantial probability to the schedule\[
\texttt{0011|0100|0110|0001},
\]which violates the no-consecutive-duty constraint and therefore lies outside the fully feasible subspace. Remarkably, its raw assignment cost, $C(q)
=
3.730372,$
is even lower than the true feasible optimum, $C_{\min}
=
5.011797.$
This example illustrates an inherent limitation of partially constraint-preserving dynamics. Although the penalty term raises the total optimization energy of infeasible schedules, their underlying assignment cost may remain extremely attractive. Consequently, part of the variational amplitude continues to accumulate on low-cost but infeasible configurations, reducing the probability assigned to feasible solutions.

By contrast, every highly probable state generated by Guarded-XY is fully feasible. Since infeasible schedules are removed from the transition graph itself, no probability can leak outside \(F_{\mathrm{full}}\). The improved distribution quality therefore arises from modifying the geometry of the variational search space rather than from increasing penalty strengths. This distinction is central to the philosophy developed in this work: constraint preservation is achieved dynamically through the mixer instead of energetically through the phase Hamiltonian.

\subsection{Structural Sensitivity of the Guarded-XY+tight Ansatz}
\label{subsec:guarded_structural_sensitivity}

The previous results considered the benchmark workload
\(
W=[2,1,2,1],
\)
for which no saturated adjacent-day segments occur and the complete mixer reduces to the local guarded exchange operator. We next investigate situations in which the additional tight-pattern component becomes active.
The key structural quantity is the workload tightness,
$N_{\mathrm{tight}}$,
defined as the number of adjacent-day pairs satisfying
$W_d+W_{d+1}=N.$
Whenever this condition holds, the no-consecutive-duty constraint forces complementary worker assignments across neighboring days, substantially restricting the connectivity of the feasible-state graph. As discussed in Sec.~2.3 and Appendix~A, the  the Guarded-XY+tight mixer supplements the local guarded exchanges with collective pattern exchanges that improve feasible-sector exploration while preserving exact feasibility.

For each value of \(N_{\mathrm{tight}}\), we select a representative workload instance and optimize the Guarded-XY+tight ansatz using CVaR-QAOA. To quantify the benefit of the variational evolution independently of the size of the feasible subspace, we define the optimal-state enrichment
\[
\eta_{\mathrm{opt}}
=
\frac{
P_{\mathrm{opt}}^{\mathrm{Guarded}}
}{
P_{\mathrm{opt}}^{\mathrm{uniform}}
},
\]
where
\(P_{\mathrm{opt}}^{\mathrm{uniform}}\)
denotes the probability obtained by uniform sampling over the fully feasible subspace. Thus,
\(
\eta_{\mathrm{opt}}>1
\)
indicates amplification of the optimal feasible schedule beyond the uniform baseline.

\begin{figure}[t]
    \centering
    \includegraphics[width=0.78\linewidth]
    {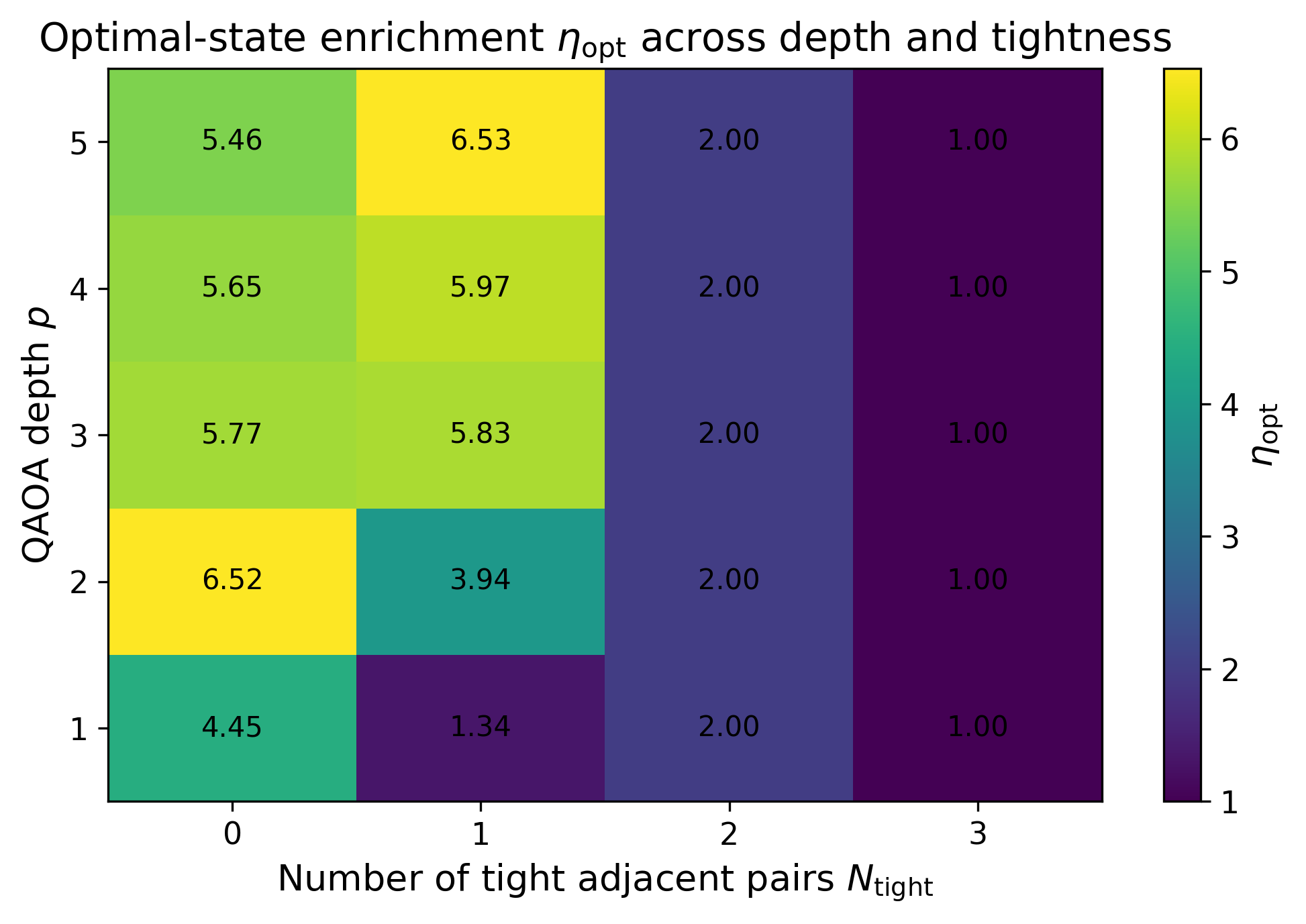}
    \caption{
    Optimal-state enrichment \(\eta_{\mathrm{opt}}\) for the Guarded-XY+tight ansatz as a function of QAOA depth and workload tightness \(N_{\mathrm{tight}}\). Each column corresponds to a representative workload with the indicated number of saturated adjacent-day pairs.
    }
    \label{fig:guarded_eta_depth_tightness}
\end{figure}

The results are shown in Fig.~\ref{fig:guarded_eta_depth_tightness}. For workloads with small values of \(N_{\mathrm{tight}}\), the enrichment generally increases with circuit depth, indicating that additional constraint-preserving layers continue to improve concentration on the optimal feasible schedule. In contrast, more saturated workloads exhibit weaker depth dependence and earlier saturation. This behaviour is consistent with the underlying graph structure: as adjacent-day complementarity becomes more restrictive, the feasible subspace contains fewer independent directions along which probability can be redistributed.

These calculations should be interpreted as representative structural sensitivity tests rather than a statistical survey over workload classes. Nevertheless, they demonstrate that the benefit of increasing circuit depth depends not only on the objective landscape but also on the connectivity of the feasible-state graph. Flexible workloads admit progressively stronger optimal-state enrichment, whereas highly saturated workloads become increasingly rigid even under a fully constraint-preserving mixer.

\subsection{Localized Feasible-State Initialization}

\begin{figure*}[t]
    \centering

    \begin{subfigure}[t]{0.49\textwidth}
        \centering
        \includegraphics[width=\linewidth]
        {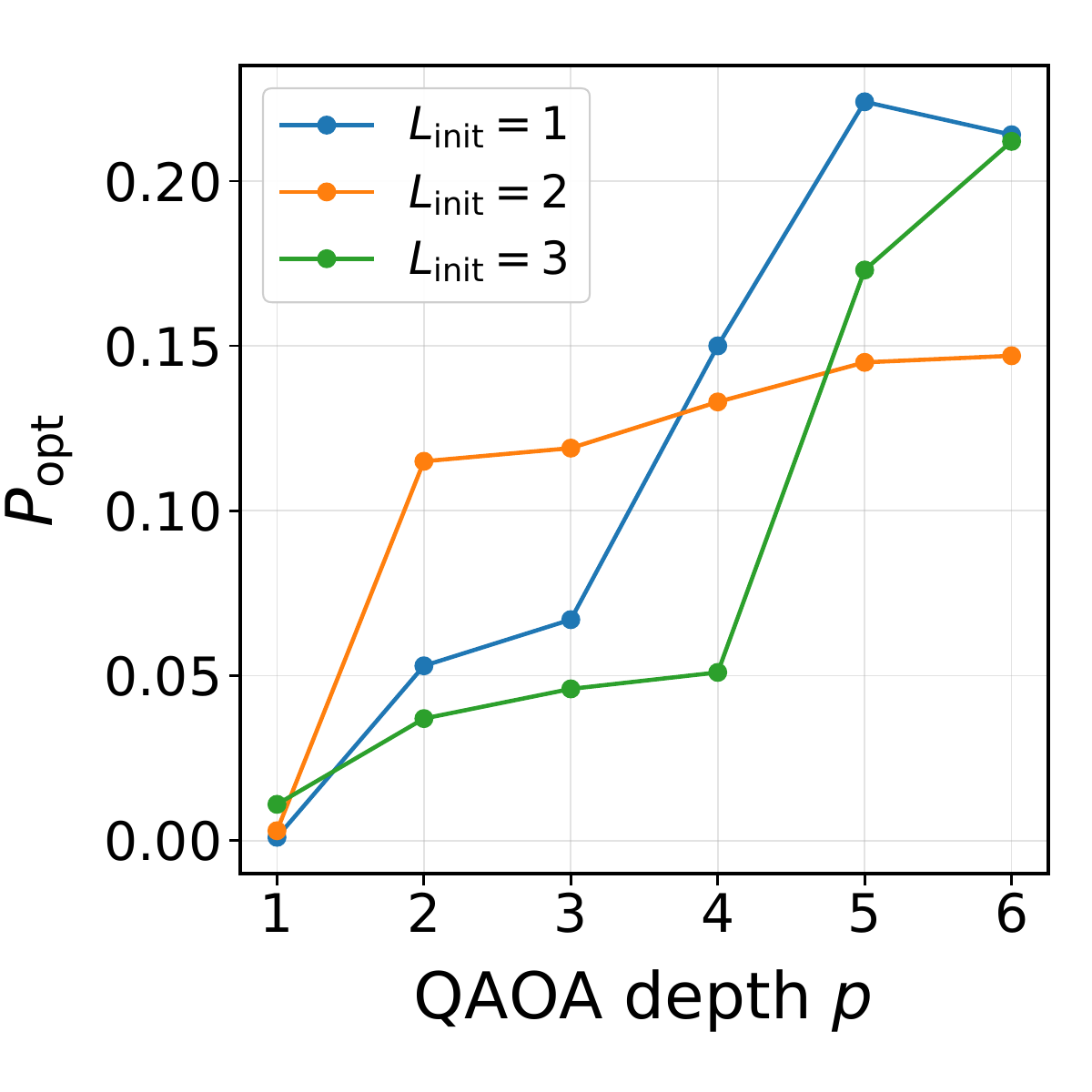}
        \caption{}
    \end{subfigure}
    \hfill
    \begin{subfigure}[t]{0.49\textwidth}
        \centering
        \includegraphics[width=\linewidth]
        {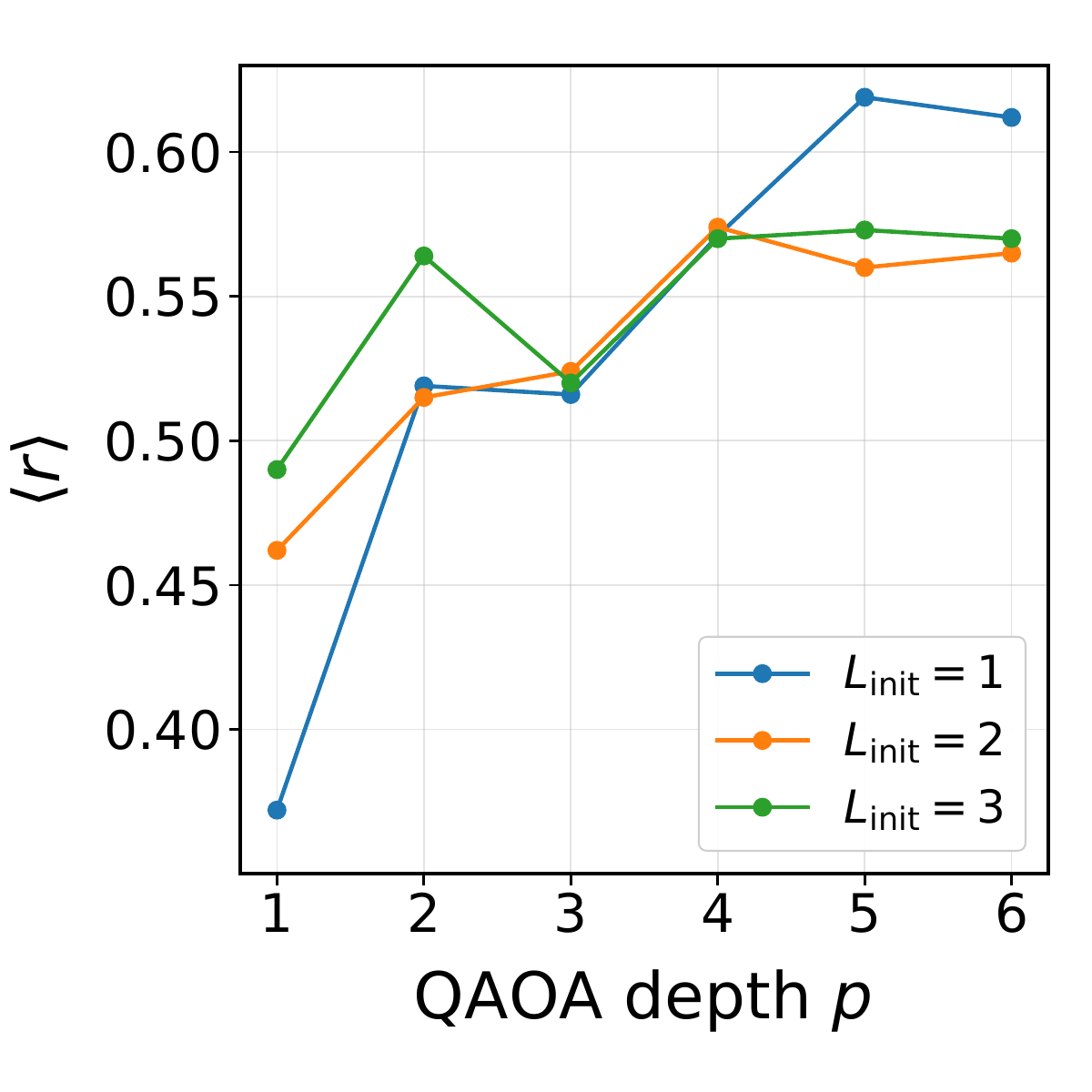}
        \caption{}
    \end{subfigure}

    \caption{
    Dependence of Guarded-XY performance on localized feasible-state initialization. Panel (a) shows the optimal-state probability $P_{\mathrm{opt}}$, while panel (b) shows the mean normalized score \(\langle r\rangle\).
    }
    \label{fig:guarded_initial_state_sensitivity}
\end{figure*}
The preceding results assume that the initial state is the uniform superposition over the fully feasible subspace. While this choice is convenient for isolating the properties of the mixer, preparing such a state may itself become challenging for larger problems. We therefore investigate whether the Guarded-XY ansatz can instead be initialized from a single feasible schedule while retaining the advantages of constraint-preserving dynamics~\cite{bartschi2019dicke}. Figure~\ref{fig:guarded_initial_state_sensitivity} shows the dependence of the optimization on the number of mixer-only spreading layers applied before the variational QAOA circuit. Since the spreading layers are generated by the same guarded mixer used during the variational optimization, the evolution remains entirely inside the fully feasible subspace. Consequently, any differences observed in Fig.~\ref{fig:guarded_initial_state_sensitivity} arise solely from the redistribution of probability within the feasible-state graph rather than from leakage into infeasible configurations.

The results show that substantial optimal-state amplification can already be achieved when starting from a single non-optimal feasible roster. A single spreading layer produces the strongest performance and reaches
$P_{\mathrm{opt}}
=
0.223555$ at intermediate circuit depth. Increasing the number of spreading layers does not necessarily improve the optimization and may instead distribute probability more uniformly over the feasible subspace, reducing concentration on the optimal schedule.

This observation has practical significance. Preparing a uniform superposition over all feasible schedules may require a nontrivial state-preparation circuit, whereas initializing a single feasible roster requires only computational-basis preparation followed by a small number of mixer applications ~\cite{egger2021warmstarting}. The localized-start strategy therefore offers a hardware-friendly alternative that retains the principal advantage of the constraint-preserving framework, namely that every stage of the evolution remains inside the fully feasible search space.

\subsection{Ground-State Probability Across Different Instance Sizes}
\label{subsec:ground_state_probability_scaling}
\begin{figure}[t]
    \centering

    \begin{subfigure}{0.48\linewidth}
        \centering
        \includegraphics[width=\linewidth]
        {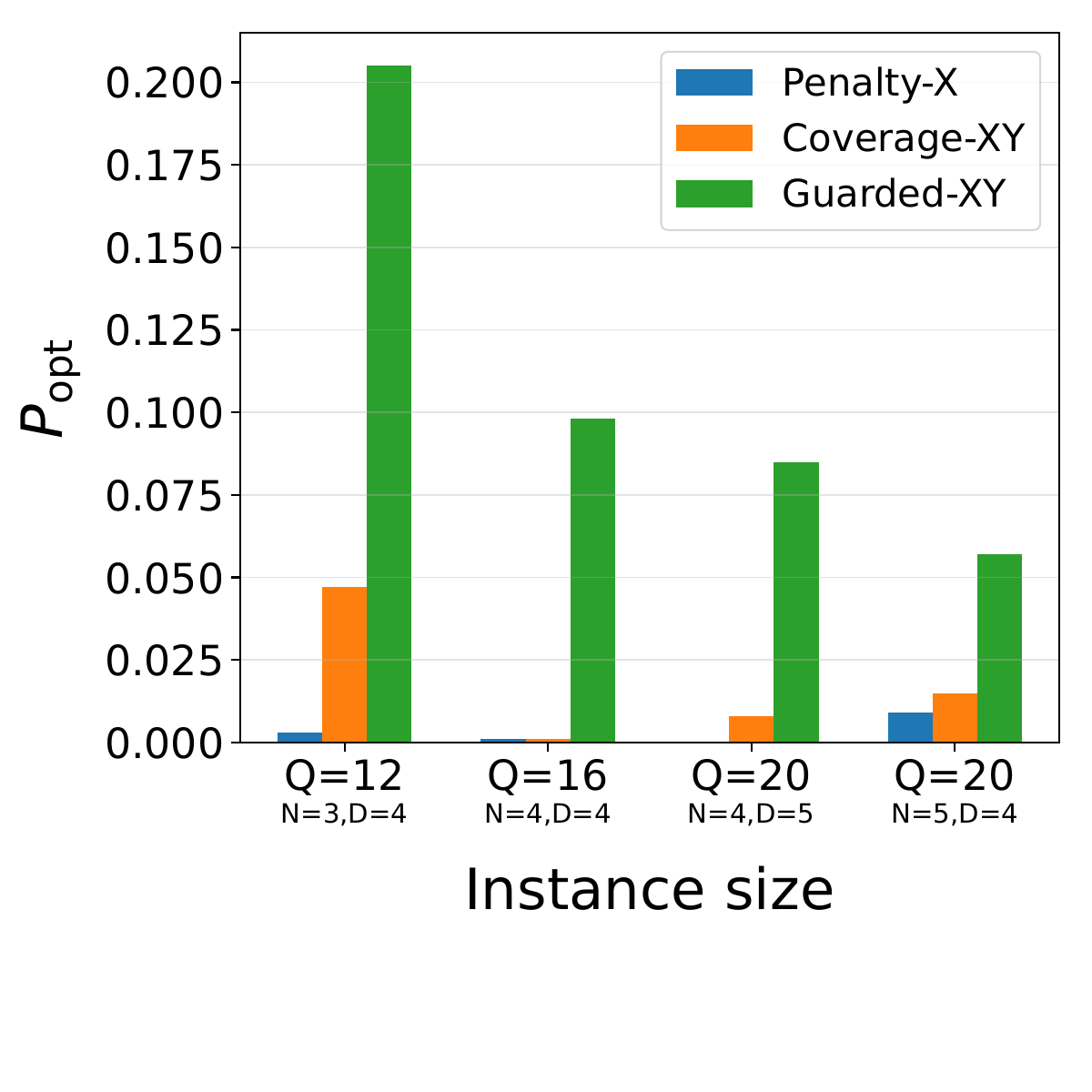}
        \caption{}
\label{fig:ground_state_probability_scaling_a}
    \end{subfigure}
    \hfill
    \begin{subfigure}{0.48\linewidth}
        \centering
        \includegraphics[width=\linewidth]
        {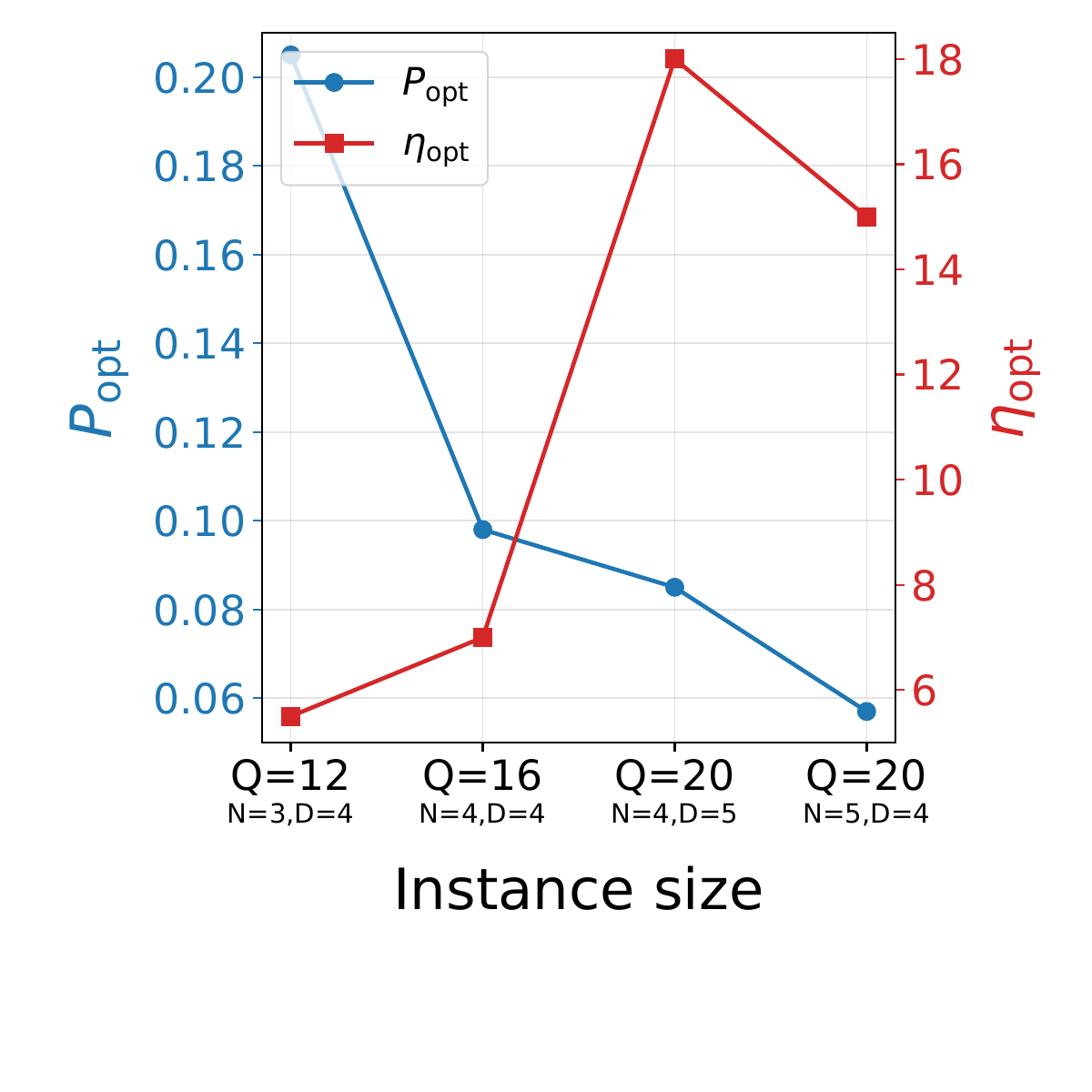}
        \caption{}
\label{fig:ground_state_probability_scaling_b}
    \end{subfigure}
    \caption{
    Fixed-depth comparison across several benchmark instances.
    (a) Ground-state probability $P_{\mathrm{opt}}$ for Penalty-X, Coverage-XY, and Guarded-XY at depth $p=1$. Guarded-XY consistently gives the largest probability of sampling the optimal feasible roster.
    (b) Guarded-XY ground-state probability together with the optimal-state enrichment
$\eta_{\mathrm{opt}}=|F_{\mathrm{full}}|P_{\mathrm{opt}}$.
    Although the raw optimal-state probability decreases for larger instances, the enrichment factor demonstrates continued amplification relative to uniform sampling over the fully feasible subspace.
    }
\label{fig:ground_state_probability_scaling}
\end{figure}

To assess whether the behaviour observed for the \(16\)-qubit benchmark is representative of other small rostering problems, we repeated the comparison on several additional instances with different numbers of workers and scheduling days~\cite{shaydulin2019evaluating,lotshaw2021empirical}.  The test set consists of one \(12\)-qubit instance, the main \(16\)-qubit benchmark, and two structurally distinct \(20\)-qubit instances.

Since the larger benchmarks were not all optimized to the same maximum circuit depth, we compare the three ans\"atze at the common available depth $p=1.$
For the diagonal cost Hamiltonians considered here, the ground state corresponds to the optimal feasible roster. Assuming a non-degenerate optimum, $P_{\mathrm{GS}}
=P_{\mathrm{opt}}.$
Figure~\ref{fig:ground_state_probability_scaling}(a) compares the three ans\"atze across the tested benchmarks. For every instance considered, Guarded-XY yields the largest probability of sampling the exact feasible optimum. For the \(12\)-qubit benchmark,
$P_{\mathrm{opt}}
=
0.204132,$
compared with $0.046710$ for Coverage-XY and $0.003183$ for Penalty-X. The same qualitative ordering is observed for the \(16\)-qubit benchmark and for both structurally different \(20\)-qubit instances. Thus, even at shallow depth, embedding both hard constraints directly into the mixer consistently increases the probability of obtaining the optimal feasible roster.

The raw optimal-state probability, however, depends on the size of the feasible subspace and therefore should not be interpreted as a complete measure of size dependence. To normalize this effect, we also consider the optimal-state enrichment
\[
\eta_{\mathrm{opt}}
=
\frac{P_{\mathrm{opt}}^{\mathrm{Guarded}}}
     {P_{\mathrm{opt}}^{\mathrm{uniform}}}
=
|F_{\mathrm{full}}|P_{\mathrm{opt}}^{\mathrm{Guarded}},
\]
where
\(
|F_{\mathrm{full}}|
\)
denotes the number of fully feasible schedules. By construction, $\eta_{\mathrm{opt}}
=
1$ corresponds to uniform sampling over the feasible subspace, while
\(
\eta_{\mathrm{opt}}>1
\)
measures amplification of the optimum beyond the uniform baseline.

The corresponding results are shown in Fig.~\ref{fig:ground_state_probability_scaling}(b). Although the raw value of
\(P_{\mathrm{opt}}\)
decreases as the benchmark instances become larger, the enrichment factor remains substantially above unity and even increases for the larger tested problems. This indicates that the reduction in raw optimal-state probability is largely a consequence of the growing feasible search space rather than a loss of optimization capability. Relative to uniform feasible sampling, the Guarded-XY ansatz continues to preferentially concentrate probability on the optimal feasible schedule across all tested benchmarks.

The present calculations should be interpreted as a small-instance sensitivity study rather than a scaling analysis. The benchmark instances differ not only in qubit number but also in workload structure and feasible-subspace size, and the two \(20\)-qubit examples correspond to different combinations of workers and scheduling days. Nevertheless, the qualitative behaviour is consistent across all tested cases. Embedding the hard constraints directly into the mixer continues to provide the highest fixed-depth optimal-state probability while maintaining substantial optimal-state enrichment relative to uniform sampling over the fully feasible subspace.

\section{Conclusion}
We have developed a constraint-preserving QAOA framework for personnel rostering in which hard scheduling constraints are incorporated directly into the quantum dynamics rather than enforced solely through energetic penalty terms~\cite{Hadfield2017HardSoftConstraints, Hadfield2019QAOAtoQAOAplus}. The central idea is to construct mixer Hamiltonians whose transition graphs coincide with progressively smaller feasible sectors of the scheduling problem. This viewpoint shifts constraint handling from the phase Hamiltonian to the mixer and provides a geometric interpretation of variational quantum optimization in terms of dynamically accessible configuration spaces.

Building on the conventional Coverage-XY construction \cite{Wang2020XYMixers}, we introduced a guarded transition operator that removes every elementary exchange capable of violating the no-consecutive-duty constraint. The resulting Guarded-XY mixer preserves both daily coverage and the no-consecutive-duty constraints exactly throughout the quantum evolution, making the fully feasible subspace an invariant subspace of the dynamics. Since infeasible schedules are never generated, the phase Hamiltonian reduces to the assignment-cost operator alone, eliminating the need for hard-constraint penalty coefficients. We further showed that exact feasibility preservation and the design of useful feasible transitions are distinct considerations. This motivated the introduction of the tight-pattern extension, which adds collective feasible exchanges in saturated workload segments where local guarded exchanges alone are insufficient.

Using exact statevector simulations, we compared Penalty-X, Coverage-XY, and Guarded-XY on representative personnel-rostering benchmarks under both expectation-value and CVaR optimization. Across the studied instances, the fully constraint-preserving construction consistently produced the highest-quality output distributions. The guarded dynamics maintained unit feasibility probability by construction while achieving strong concentration on the optimal feasible roster. The numerical study also demonstrated that the framework remains effective when initialized from localized feasible schedules and that its advantage persists across several benchmark instances with different workload structures.

Personnel rostering has previously been formulated for quantum annealing and other Ising-based optimization approaches, where hard constraints are typically enforced through carefully tuned penalty Hamiltonians~\cite{Ikeda2019NurseQA}. In contrast, the present work develops, to the best of our knowledge, the first systematic constraint-preserving QAOA formulation for personnel rostering in which both hard scheduling constraints are embedded directly into the mixer Hamiltonian. The resulting variational evolution is therefore confined to the feasible scheduling subspace from the outset rather than relying on energetic suppression of infeasible configurations.

Although the numerical demonstrations presented here are limited to small benchmark instances accessible by exact simulation, the underlying construction is independent of system size. The guarded transition principle depends only on local scheduling constraints, while the separation between feasibility preservation and feasible-transition design provides a general recipe for designing mixers for constrained combinatorial optimization problems. Beyond personnel rostering, the same methodology should be applicable to a broad class of scheduling, assignment, routing, and resource-allocation problems in which feasible configurations can be characterized through local transition rules. We hope that this dynamical perspective on constraint preservation will provide a useful framework for the development of future QAOA algorithms for constrained optimization.
\appendix

\section{Tight-pattern guarded mixer}
\label{app:tight_pattern_mixer}

This appendix gives the operator-level construction of the tight-pattern contribution used in the Guarded-XY+tight mixer. The purpose of this term is to supplement the local one-day guarded exchanges with collective feasible exchanges in saturated workload segments, while preserving both hard scheduling constraints exactly.

Consider two adjacent days \(d\) and \(d+1\). If their workloads satisfy
\begin{equation}
W_d+W_{d+1}=N,
\label{eq:saturated_condition_appendix}
\end{equation}
then the no-consecutive-duty constraint forces the assignments on the two days to be complementary: every worker assigned on day \(d\) must be unassigned on day \(d+1\), and every worker unassigned on day \(d\) must be assigned on day \(d+1\). More generally, a maximal saturated segment is a consecutive set of days
\[
C=\{a,a+1,\ldots,b\}
\]
over which neighboring workload pairs satisfy Eq.~\eqref{eq:saturated_condition_appendix}. Let \(\mathcal T\) denote the set of all such maximal saturated segments.

Within a saturated segment, local one-day guarded exchanges may be insufficient to change between feasible alternating patterns. We therefore introduce a collective exchange between two workers \(u\) and \(v\) across the whole saturated segment. Let
\[
s^{(0)}=(s^{(0)}_0,s^{(0)}_1,\ldots,s^{(0)}_{|C|-1})
\]
denote one alternating binary pattern on the segment, and let
\[
s^{(1)}=1-s^{(0)}
\]
denote its complement. The collective exchange swaps the two complementary patterns between workers \(u\) and \(v\).

Using
\[
\sigma^\pm_{n,d}
=
\frac{X_{n,d}\pm iY_{n,d}}{2},
\]
define the elementary pattern-changing operator
\[
B^{(0)}_{u,v,d}
=
\sigma^+_{u,d}\sigma^-_{v,d},
\qquad
B^{(1)}_{u,v,d}
=
\sigma^-_{u,d}\sigma^+_{v,d}.
\]
The collective alternating-pattern exchange operator over the segment \(C=\{a,\ldots,b\}\) is then
\begin{equation}
\hat A^C_{u,v}
=
\prod_{j=0}^{|C|-1}
B^{(s^{(0)}_j)}_{u,v,a+j}.
\label{eq:appendix_collective_exchange}
\end{equation}
This operator maps the configuration in which worker \(u\) carries pattern \(s^{(1)}\) and worker \(v\) carries pattern \(s^{(0)}\) to the configuration in which their patterns are exchanged. The Hermitian conjugate \(\hat A^{C\dagger}_{u,v}\) generates the reverse exchange.

The collective exchange preserves the daily coverage constraint because, on every day in the segment, one occupation is removed and one occupation is added. It also preserves the internal no-consecutive-duty constraint within the saturated segment because the exchanged worker patterns remain complementary. Therefore, the only possible new violations can occur at the segment boundaries, namely across the pairs \((a-1,a)\) and \((b,b+1)\), if those days exist.

To prevent such boundary violations, we introduce a boundary guard projector \(\hat G^C_{u,v}\). This projector removes collective exchanges that would create a no-consecutive-duty violation across either boundary of the saturated segment. Equivalently, \(\hat G^C_{u,v}\) excludes boundary configurations for which the post-exchange occupations of workers \(u\) or \(v\) would create
\[
q_{n,d}q_{n,d+1}=1
\]
across \(a-1,a\) or \(b,b+1\). The tight-pattern mixer contribution is therefore
\begin{equation}
\hat H_M^{\mathrm{tight}}
=
\sum_{C\in\mathcal T}
\sum_{u<v}
\hat G^C_{u,v}
\left(
\hat A^C_{u,v}
+
\hat A^{C\dagger}_{u,v}
\right).
\label{eq:appendix_tight_mixer}
\end{equation}

Combining the one-day guarded exchange term of Eq.~\eqref{eq:guarded_mixer} with the tight-pattern contribution gives the Guarded-XY+tight mixer,
\begin{equation}
\hat H_M^{\mathrm{GXT}}
=
\sum_{d=1}^{D}
\sum_{u<v}
\hat G_{u,v,d}
\left(
\sigma^+_{u,d}\sigma^-_{v,d}
+
\sigma^-_{u,d}\sigma^+_{v,d}
\right)
+
\sum_{C\in\mathcal T}
\sum_{u<v}
\hat G^C_{u,v}
\left(
\hat A^C_{u,v}
+
\hat A^{C\dagger}_{u,v}
\right).
\label{eq:appendix_gxt_mixer}
\end{equation}

The first term in Eq.~\eqref{eq:appendix_gxt_mixer} is the local guarded mixer defined in Sec.~\ref{subsec:guarded_transition_operator}. The second term adds collective feasible exchanges in saturated workload segments. Both terms preserve daily coverage, and both are guarded against no-consecutive-duty violations. Hence, when initialized in \(F_{\mathrm{full}}\), the Guarded-XY+tight mixer generates dynamics confined to the fully feasible sector.
\section*{Acknowledgements}
Aruna Gupta and S. R. Hassan acknowledge support from the QC BRIDGE Post-Doctoral Fellowship, funded by the Foundation for QC Innovation (FQCI), DST-NQM Hub for Quantum Computing at the Indian Institute of Science, Bengaluru.
% References are embedded directly so the journal processor does not need to run BibTeX.

\end{document}